%% file: main.tex
\documentclass[lettersize,journal]{./template-files/IEEEtran}
\usepackage{cite}
\usepackage{ifpdf}
\usepackage{xurl}
\usepackage{array}
\usepackage{mdwmath}
\usepackage{fixltx2e}

\input{misc/packages}
\input{misc/config}
\input{misc/macros}

\usepackage{tikz}
\usetikzlibrary{automata,arrows,positioning,calc}
\usepackage{etoolbox}
\usepackage{bbm}
\makeatletter
\patchcmd{\@makecaption}
  {\scshape}
  {}
  {}
  {}
\makeatother

\begin{document}

\title{CATS: A framework for Cooperative Autonomy\\Trust \& Security}

\author{Namo Asavisanu, 
Tina Khezresmaeilzadeh,
Rohan Sequeira,
Hang Qiu,\\
Fawad Ahmad,
Konstantinos Psounis,
and Ramesh Govindan

\thanks{© 20XX IEEE.  Personal use of this material is permitted.  Permission from IEEE must be obtained for all other uses, in any current or future media, including reprinting/republishing this material for advertising or promotional purposes, creating new collective works, for resale or redistribution to servers or lists, or reuse of any copyrighted component of this work in other works.}
\thanks{N. Asavisanu, T. Khezresmaeilzadeh, R. Sequeira, K. Psounis, and R. Govindan are with the University of Southern California, Los Angeles, CA, USA. (e-mail: \{namo, khezresm, rsequeir, kpsounis, ramesh\}@usc.edu)}%
\thanks{H. Qiu is with the University of California, Riverside, Riverside, CA, USA. (e-mail: hangq@ucr.edu)}%
\thanks{F. Ahmad is with the Rochester Institute of Technology, Rochester, NY, USA. (e-mail: fawad@cs.rit.edu)}%
}%

\maketitle

\input{sections/abstract}

\input{sections/introduction}

\input{sections/adversary-model}

\input{sections/system-design}

\input{sections/experimental-evaluation}

\input{sections/risk-analysis-new}

\input{sections/related-works}

\input{sections/conclusion}

\input{sections/acknowledgement}

\appendices
\input{sections/appendix-parameter-selection.tex}

\bibliographystyle{ieeetr}

\begingroup
    \bibliography{references/ref.bib}
\endgroup

\end{document}

%% file: misc/packages.tex
\usepackage[a-1b]{pdfx}

\usepackage{amsmath,amsfonts}
\usepackage[T1]{fontenc}
\usepackage[utf8]{inputenc}

\usepackage{upquote}

\usepackage{microtype}
\UseMicrotypeSet[protrusion]{basicmath} 

\usepackage{svg}

\usepackage{hyperref}
\usepackage{graphicx,grffile}
\usepackage{times} 
\usepackage{algorithm, algpseudocode}
\usepackage{eqparbox}
\usepackage{multirow}
\usepackage{multicol}
\usepackage{xspace}
\usepackage{tabularx}
\usepackage{ragged2e}
\usepackage{booktabs}
\usepackage{paralist}
\usepackage[american]{babel}
\usepackage{courier,color,wrapfig}
\usepackage{balance}
\usepackage{epstopdf}
\usepackage{booktabs}
\usepackage{url}
\usepackage{pifont}
\usepackage{tikz}
\usepackage{mathtools}
\usepackage[normalem]{ulem}
\usepackage{comment}
\usepackage{changepage}

\usepackage{subcaption}

\let\labelindent\relax
\usepackage{enumitem}

\usepackage{etoolbox}
\usepackage{ifthen}
\usepackage[textsize=small]{todonotes}
\usepackage{tikz}
\usetikzlibrary[automata,calc,intersections,through,backgrounds,arrows,positioning,decorations.pathmorphing,fit]
\usepackage[compact]{titlesec}

\usepackage{xurl}

\usepackage{tikz}
\usetikzlibrary{automata,arrows,positioning,calc}
\usepackage{etoolbox}
\usepackage{bbm}
\makeatletter
\patchcmd{\@makecaption}
  {\scshape}
  {}
  {}
  {}
\makeatother


%% file: misc/config.tex
\makeatletter
\def\maxwidth{\ifdim\Gin@nat@width>\linewidth\linewidth\else\Gin@nat@width\fi}
\def\maxheight{\ifdim\Gin@nat@height>\textheight\textheight\else\Gin@nat@height\fi}
\makeatother
\setkeys{Gin}{width=\maxwidth,height=\maxheight,keepaspectratio}
\makeatletter
\g@addto@macro{\UrlBreaks}{\UrlOrds}
\makeatother

\hypersetup{nolinks=true}



\newcommand\paraspace{\vspace*{0.5ex}}
\providecommand\parab[1]{\paraspace\noindent{\textbf{#1}}}
\providecommand\parae[1]{\paraspace\noindent{\textit{#1}}}

\apptocmd\normalsize{%
\abovedisplayskip=5pt
\abovedisplayshortskip=5pt
\belowdisplayskip=5pt
\belowdisplayshortskip=5pt
}{}{}

\algnewcommand\INPUT{\item[\textbf{Input:}]}%
\algnewcommand\OUTPUT{\item[\textbf{Output:}]}%

\savebox\strutbox{$\vphantom{\dfrac10}$}


%% file: misc/macros.tex
\newcommand{\sysname}{\textsc{CATS}\xspace}

\newcommand{\TruPercept}{\textit{TruPercept}\xspace}
\newcommand{\VR}{\textit{VANET Reputation}\xspace}

\newcommand{\revise}[1]{#1}
\newcommand{\revisetwo}[1]{#1}

\DeclareMathDelimiter{(}{\mathopen} {operators}{"28}{largesymbols}{"00}
\DeclareMathDelimiter{)}{\mathclose}{operators}{"29}{largesymbols}{"01}

\newcommand{\floatingensure}[1]{\State \textbf{Ensure:} {#1}}

\newcommand{\ie}{\textit{i.e.,}\xspace}
\newcommand{\eg}{\textit{e.g.,}\xspace}

\newcommand{\secref}[1]{\S\ref{#1}}
\newcommand{\figref}[1]{Figure~\ref{#1}}
\newcommand{\tabref}[1]{Table~\ref{#1}}
\newcommand{\algoref}[1]{Algorithm~\ref{#1}}
\newcommand{\eqnref}[1]{Equation~\ref{#1}}

\newcommand{\apdxref}[1]{Appendix~\ref{#1}}

\newcommand*\circled[1]{\tikz[baseline=(char.base)]{
              \node[shape=circle,draw,fill=gray!30,inner sep=1pt] (char) {#1};}}


%% file: sections/abstract.tex
\begin{abstract}
With cooperative perception, autonomous vehicles can wirelessly share sensor data and representations to overcome sensor occlusions, improving situational awareness. Securing such data exchanges is crucial for connected autonomous vehicles. Existing, automated \textit{reputation-based} approaches often suffer from a delay between detection and exclusion of misbehaving vehicles, while \textit{majority-based} approaches have communication overheads that limits scalability. 
In this paper, we introduce \sysname, a novel automated system that blends together the best traits of reputation-based and majority-based detection mechanisms to secure vehicle-to-everything (V2X) communications for cooperative perception, while preserving the privacy of cooperating vehicles.
Our evaluation with city-scale simulations on realistic traffic data shows \sysname's effectiveness in rapidly identifying and isolating misbehaving vehicles, with a low false negative rate and overheads, proving its suitability for real world deployments.
\end{abstract}

%% file: sections/introduction.tex
\section{Introduction}
\label{sec:introduction}

Recent years have seen tremendous improvements in autonomous driving technologies, pushing the deployment of self-driving cars closer to commercial realization~\cite{waymo-cpuc}. As deployments scale, more challenging corner cases surface to stress-test the reliability of the autonomous driving system. Examples of these corner cases include limited visibility due to occlusion, degraded perception at long range, transient reflection and so on. In order for self-driving cars to be widely acceptable, they need to be able to tackle these corner cases.

To address limited visibility, recent research on cooperative perception~\cite{AVR, autocast, ren2023interruptionaware, aoki2020cooperative, kim2015impact, chen2019fcooper, emp, shi2022vips, he2021vi, he2022automatch, BEVDataSharing} proposes to leverage V2X communication to share perception data with nearby vehicles to fill in the invisible area. 
Cooperative perception enhances the performance of onboard tasks such as detection~\cite{v2vnet, v2xvit}, motion prediction~\cite{v2vnet,WangIntentSharing,cooperv2x}, real-time mapping~\cite{xu2022cobevt}, planning and control~\cite{coopernaut}.

While the benefit of cooperative perception is encouraging, the underlying assumption is that the cooperating agent is \textit{trustworthy}.
In other words, the data it shares is always authentic and valid. However, it is very challenging yet crucial to implement \textit{trust} in the data exchange, as recent literature \cite{MultiSourceAdversarialAttack,securingv2x} has illustrated multiple methods that a malicious attacker could use to attack a V2X cooperative network.

\begin{figure}
    \centering
    \includegraphics[width=6.5cm]{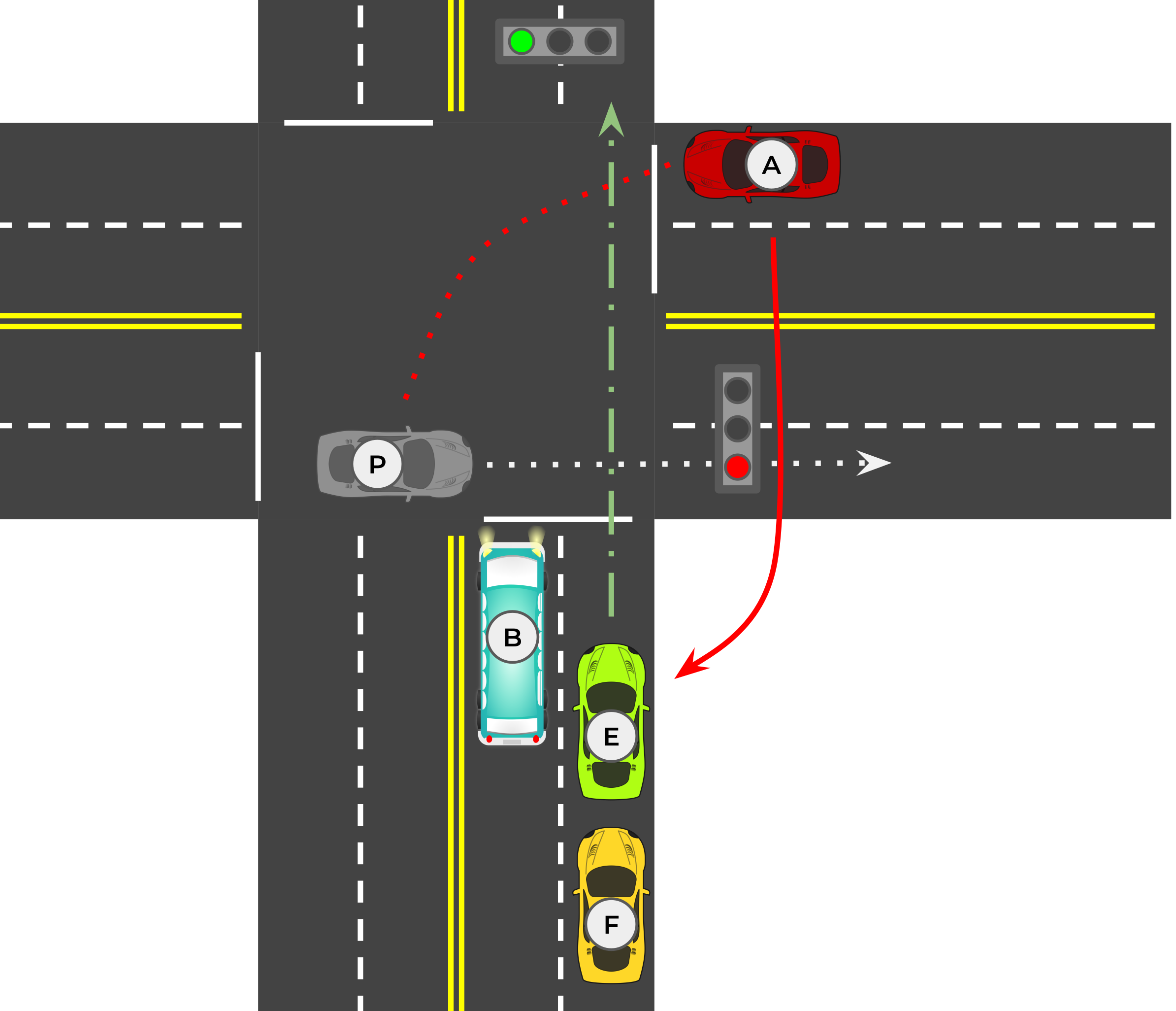}
    \caption{Phantom red light violator: a \textit{data injection} attack.}
    \label{fig:red-light-violator-scenario}
\end{figure}

To demonstrate a specific case, we stage a proof of concept attack in a cooperative perception~\cite{autocast} scenario (\figref{fig:red-light-violator-scenario}). In this scenario, $E$ is driving straight through the intersection. Due to the occlusion from $B$ on the left, $E$ is communicating with $B$ to provide blind spot visibility. An attacker ($A$) can falsify the presence of a non-existent \textit{phantom car} ($P$) running a red light in front of $B$. Misguided by the fake existence of the phantom car, $E$ would slow down or stop to avoid the collision, resulting in traffic backlog or a rear-end hazard.

In this paper, we ask the question of "how can we guarantee trusted data sharing in the presence of (a) malicious actors spreading bad messages, and (b) malfunctioning sensors (e.g., out-of-calibration LiDARs) sharing faulty data?"

\parab{Existing approaches.}
Recent work \cite{autosec2022_tee,ifal} leverages \textit{trusted identities} to provide authenticity and non-repudiation guarantees on data exchanged between vehicles. In some of these systems, misbehavior detection is done by having a traceable identifier, such that the authorities will be able to manually review previous V2X data logs to investigate a misbehaving vehicle and take action. The process is laborious, and can incur a significant time to response penalty. During this period, a vehicle with faulty sensors, or a malicious attacker could transmit bad data to numerous vehicles, causing performance degradation, if not accidents, before they can be excluded from the network.

To enable autonomous responses, an alternative approach is to globally aggregate reputation of vehicles, either by individual votes~\cite{VanetReputation,TFDD} or active majority visibility~\cite{TruPercept}. However, pure reputation-based systems often suffer a misbehavior response delay, and active majority-based systems often have communication overheads with limited long-term memory that constrain scalability and performance.

\parab{Design goals.} We envision a security framework for a cooperative perception environment that eliminates \textit{misinformation}, whether that be from faulty vehicles or malicious attackers, while at the same time maximizing the utilization of cooperative perception information to allow for the significant safety benefits it provides. Additionally, we design our framework such that it is practical, deployable in the near future, imposes minimal requirements on vehicles and infrastructure, and mitigates privacy concerns over the sharing of location data.

\parab{Contributions.} In this paper, we discuss the design and implementation of \sysname, a system that enables secure and trusted cooperative autonomy. By blending the best elements of long-term, reputational voting {(long-term trust)} with {\textit{passive}} majority view {(utilizing cooperative perception data that is already transmitted in the local vicinity)} {and a multiple-party voting system that draws conclusions from diverse voters}, we created a {comprehensive and practical} filtering system that closes the gap on response times while maintaining a low communication overhead {and risk for denial of service attacks}.

\noindent In summary, \sysname provides:
\begin{itemize}[leftmargin=*]
        \item A novel, hybrid solution between reputation-based and majority-based systems for determining vehicle trust, providing \textit{the best of both worlds}.
        \item {A novel voting scheme that combines multiple reports from independent vehicles to draw a reliable conclusion, without requiring trusted onboard clocks or execution environments}.
        \item A signature and pseudonym method agnostic solution that is adaptable and future-proof to keep up with latest threats.
        \item A method to exchange pseudonyms and trust data in a peer-to-peer fashion \revise{without the need for road-side units (RSUs)}, enabling near-future deployment without extensive investment in additional infrastructure.
\end{itemize}

We implemented \sysname on top of SUMO\cite{sumo2018}, a city-scale simulator. Our evaluation results in two cities (\secref{sec:experimental-evaluation}) shows that \sysname is extremely effective in filtering out {bad messages from misbehaving vehicles, with an average 230x reduction of bad messages, while making a small (4.2x on average) tradeoff for blocking good messages, when compared to other approaches.}

We also performed a formal analysis (\secref{sec:risk-analysis}) to {extrapolate long-term performance expectations of \sysname, and show that our theoretical model lines up with experimental results.}

%% file: sections/adversary-model.tex
\section{Adversary Model \& Assumptions}

In this paper, we primarily focus on preventing bad data {(\ie inaccurate data)} of \textit{misbehaving} vehicles from affecting others' control decisions. This includes \textit{malfunctioning} vehicles, which can unintentionally share bad data as a result of a malfunctioning sensor, and \textit{malicious} vehicles, which deliberately share bad data. In the latter case, we focus on adversaries who would try to bypass the protection given by \sysname through influencing the system's own mechanisms {(\ie through manipulation of reputation)}. We {also} make the following {additional} assumptions:

    \parae{Vehicles}: An adversary may have full software and hardware access to vehicles that they own. However, we assume that a vehicles' individual private keys are securely stored in tamper-proof hardware and cannot be extracted. 

    \parae{Sensor attacks}: Sensor-based attacks (e.g., fooling LiDARs with lasers) are considered out of scope (see \cite{sato2024lidar} and references therein on methods, defenses, and limitations of such attacks).
        
    \parae{Communication}: Two communication methods are available to the vehicles at all times: the \textit{in-band, delay-sensitive} V2X network (which contains multiple independent \revise{broadcast} channels), and the internet (considered out-of-band and delay-tolerant). The V2X channel by itself does not guarantee security nor privacy, but the out-of-band communication channel for voting and trust synchronization is less delay sensitive and is secured with standard measures (\eg TLS). Communications jamming is out of scope.
    
    \parae{Identity}: Identities are securely provisioned and is immutable, being protected by secure public-key cryptography.
    
    \parae{Authority}: The centralized Security Authority (SA) is trusted, secured, and expected to maintain privacy of vehicles by not revealing the link between vehicle identities and their pseudonyms {or attempting to infer travel history from votes.} \revise{Out-of-band attacks on the SA itself is out of scope.}

%% file: sections/system-design.tex
\section{System Design}
\label{sec:systemdesign}

\subsection{System overview}
Suppose that we have autonomous vehicles with cooperative perception, using \sysname to enhance trust and security. As the vehicles drive around, cooperative perception messages are broadcasted to surrounding vehicles through V2X channels.

\sysname's {onboard component}, installed on every vehicle, verifies authenticity, integrity, {freshness,} and trustworthiness of messages in real time. The voting system, installed on every vehicle, evaluates nearby vehicles' {reliability and} trustworthiness, before submitting results to a centralized authority, enabling \textit{peer review} based trust aggregation over time.

\input{figure_latex/system}

\figref{fig:sysdesign-vehicle} illustrates how \sysname cooperatively determines trust. First, a transmitting vehicle sends cryptographically signed cooperative perception messages \circled{1}. On each receiving vehicle, a data verification module \circled{2} {checks} the validity of the message that was received, using reference data from their own sensors.

Based on this output from the {verification} system, vehicles can create {and submit a \textit{vote} \circled{3} to} the centralized SA through a voting channel (\eg the internet) \circled{4}.

By utilizing vote results from multiple vehicles, the SA can etermine the {trustworthiness} of each vehicle in the network. Vehicles that share inaccurate data, whether because of a faulty sensor, or a deliberate attack, {can then be barred by the SA from participating in the cooperative perception network} \circled{5}.

The SA then creates and signs trust updates based on latest trust information \circled{6}, which is then pushed to vehicles \circled{7}. This information is used by all participants to filter out and drop untrustworthy messages from their perception pipeline.

A vehicle will only make use of cooperative perception data from another vehicle if the sender vehicle is in a \textit{trusted} state. For sender vehicles in untrusted states, the receiving vehicle will not incorporate the received data into its perception pipeline, but will still run it through the {verification} system, helping untrusted vehicles rebuild their own {reputation} over time.

\input{figure_latex/arch}

In the following subsections, we describe each component of \sysname in more detail. Figure \ref{fig:cavsec-system-components} visualizes how all components of \sysname fit together to create a complete system.

\subsection{Voting mechanism}
\label{sec:vote_mechanism}

\revise{When a cooperative perception message is received by a vehicle, a verification system, such as \cite{CADConsistencyChecking}, cross-checks} whether the received observations are consistent with its own observations using on-board sensors, as long as there is shared visibility. A vehicle can then proceed in one of 3 ways:

\begin{itemize}[leftmargin=*]
    \item \textbf{Confirmed valid data $\rightarrow$ Up-vote:} If the {observations} received through cooperative perception can be positively verified by our own sensors, we \textit{up-vote}.
    \item \textbf{Confirmed inconsistent data $\rightarrow$ Down-vote:} If the {observations} received through cooperative perception can be confirmed as \text{inconsistent} with our own sensors {(e.g., we received data for a purported object that is purportedly directly in front of us with no occlusion, but we can confirm is not there)}, we \textit{down-vote}.
    \item \textbf{Unconfirmed data:} {If the observations received through cooperative perception \textit{cannot} be confirmed as either \textit{consistent} or \textit{inconsistent} with our own sensors, we do not perform any voting actions to preserve fairness.}
\end{itemize}

\revise{As votes from a vehicle are not considered reliable, \sysname is designed to ensure} that long term reputation-affecting votes come from a diverse viewpoint, \revise{and that misbehaving voters are prevented from colluding} to artificially inflate a vehicle's reputation or perform a denial-of-service attack \revise{at scale. To do so, we introduce two rate limits:} first, a general vote limit (inter-vote epoch, $T_{IVE})$ only lets a vehicle vote for another specific vehicle once per epoch (to ensure voting diversity). Another concurrently active downvote limit (inter-downvote epoch, $T_{IDE}$) only lets a vehicle downvote a single other vehicle per \revise{its} epoch (to prevent denial-of-service, \revise{$T_{IDE}$ is multiplicatively scaled on a per-vehicle basis upon detection of subsequent downvotes to the same vehicle in short succession}).

\revise{These two rate limits, when combined with how \sysname processes votes (\secref{sec:vote_processing}), ensures that a malicious actor will need two vehicles to perform one attack, after which those two vehicles cannot be used for an attack for the limit duration, rendering attacks unsustainable (we discuss this more in \secref{sec:preventing-collusions}).}

In order for \sysname to continuously set the optimal value for $T_{IVE}$, the SA can measure statistics on timing for accepted votes, then extrapolate to estimate the expected time for a vehicle to gain full reputation (when starting from lowest reputation), following the argument that ideally, each of these votes in the vehicle's journey to fully trusted should be unique.
Similarly, $T_{IDE}$ can also be continuously computed \revise{with probabilistic methods} based on traffic density data, \revise{intuitively to ensure that misbehaving vehicles are downvoted, while minimizing the practicality of a denial-of-service attack.}
In \secref{sec:large-scale-experiment} we show through experimentation that these limits {serve a practical purpose}, \revise{in \secref{sec:tide_scaling} we show that $T_{IDE}$ scaling on a per-vehicle level effectively prevents denial-of-service attacks}, and in \apdxref{appendix:param-selection} we further explain in detail the online learning methods for $T_{IVE}$ and \revise{the base value for} $T_{IDE}$, following the intuitions earlier mentioned.

\parab{Voting Algorithm. }
{Once a vehicle has decided to perform a voting action, it creates a signed \textit{voting ballot packet}, containing the latest beacon of the target vehicle (whom we are voting for) at the time, reason for the vote, and the action (upvote or downvote). The ballot is then submitted through the out-of-band voting network (\eg cellular connection).}

To minimize the risk of replay attacks, the SA uses a vote freshness limit $T_{vote}$ (continuously updated based on the beacon broadcast interval $T_{BBI}$ and 95th percentile measured network latency across all vehicles) to discard stale votes.

{To guard against malicious vehicles randomly voting for vehicles outside its immediate vicinity, without vehicles having to directly share location data with the SA (as that may cause privacy concerns), the target vehicle's beacon is a requirement for the voting ballot. This helps the SA ensure, with reasonable certainty, that the voter is actually in range of the target vehicle.} 

{Note that while the SA does make use of the target vehicle's beacon timestamp, which may be considered untrusted, the target vehicle themselves are incentivized by \sysname's design to be truthful and accurate, as otherwise their messages will be dropped, as we will discuss in \secref{sec:verifying-data}. This prevents an attack in which a malicious vehicle sends false timestamps in their beacons, in order to prevent downvotes to itself.}

\subsection{Trust \& vote processing}
\label{sec:vote_processing}

\parab{Security authority. } {The SA is the main trusted component of \sysname, responsible for processing all votes, calculating trust states for every vehicle, and issuing certificates to vehicles. 

\revise{In the current design of \sysname, we are not focusing on the SA itself, and we assume it to be ultimately trusted. We envision that the SA can be operated by either a national government agency, a state-level government agency, or even a private entity or consortium, and may even use decentralization methods such as \cite{SuBlockchain,PPBlockchain} to further improve on robustness.}

\figref{fig:cavsec-system-components} describes the \revise{SA's internal workings} in more detail.}

\parab{Trust states. } {In a \sysname-protected cooperative autonomy network, a vehicle will have one trust state out of 3 possible states, assigned by the SA based on aggregated vote information:}

\begin{itemize}[leftmargin=*]

    \item \textbf{Trusted:} Vehicles are \textit{full} members of the network. They can share data with other vehicles to be incorporated into control decisions, and participate in the voting process.

    \item \textbf{Untrusted:} Vehicles are \textit{probationary} members of the network. They can share data with other vehicles, but the data will only be used to validate the sender vehicle's accuracy and truthfulness, not for control decisions. Additionally, the vehicle will not be able to participate in the voting process.

    \item \textbf{Banned:} Vehicles are \textit{non-members} of the network, and can neither share data nor participate in the voting process.
\end{itemize}

The key idea is that vehicles may transition between the \textit{trusted} and \textit{untrusted} states quickly if their trustworthiness is questioned, (allowing for fast, yet robust misbehavior response) but will only be \textit{banned} if there is a strong indication from multiple voters across a time span, that a vehicle is consistently misbehaving, to achieve a balance between speed of misbehavior response and maximizing availability (by preventing nuisance bans). We discuss this more in \secref{sec:vote_processing} and \secref{sec:3states_ablation}.

\parab{Vote processing. } {To process votes, the SA performs checks (outlined in \algoref{alg:sa-vote-acceptance}) to ensure that the votes are made with valid and trusted identities (lines 1-2), fresh (lines 3 to 5), and that rate limits $T_{IVE}$ and $T_{IDE}$ (as discussed previously in \secref{sec:vote_mechanism}) are applied to prevent denial-of-service attacks and score inflation (lines 6 to 18).}
To implement these rate limits, the SA maintains a global, time-indexed lists of recently accepted upvotes (upvote list, $UL$) and downvotes (downvote list, $DL$), containing pairs of vehicles that have voted for another before.

Furthermore, the algorithm instantaneously transition vehicles from \textit{trusted} and \textit{untrusted} states depending on their reputation, using a threshold $N_{thresh}$, a parameter that controls \sysname's quick-reaction sensitivity (lines 19 to 23).
The goal of the SA is to set $N_{thresh}$ to a value that will balance quick exclusion of potentially misbehaving vehicles (higher $N_{thresh}$) with robustness towards misbehaving voters (lower $N_{thresh}$, allowing for more \textit{buffer zone} that lets vehicles lose some trust before it becomes untrusted). We further discuss our empirical derivation for a suitable value of $N_{thresh}$ in \apdxref{appendix:param-selection}.

Finally, note that the vote acceptance procedure will not make any changes to a vehicle's trust state if it is already banned, nor does it ban vehicles on its own.

\input{algorithms/sa-vote-processing.tex}

\parab{Flagging mechanism. }
To make \sysname more robust, a flagging system, instead of raw reputation, is used to decide whether a vehicle should be banned. Ideally, we would apply a ban only to vehicles with a clear indication of continuous misbehavior, both benign (malfunctioning sensor) and malicious (active attacker).

Additionally, the flagging system protects honest vehicles against the rare case where a misbehaving \textit{voter} (\eg vehicles with bad sensors or a malicious attacker) may perform a false vote, by requiring multiple reports before any action is taken.

We first \textit{discretize} the timeline into slots that we call flagging windows (each of length $T_{FW}$). During each flagging window, if there is at least one downvote, the offending vehicle is given a \textit{flag}. The type of flag assigned depends on whether said vehicle already has a flag or not - if the vehicle \revise{does not have any flags already}, a \textit{yellow} (warning) flag is given. Otherwise, a \textit{red} flag is given, and the vehicle then becomes banned. \revise{(Note that the flags does not correspond to the type of misbehavior, which \sysname does not distinguish, but rather whether the vehicle has been involved in a prior misbehavior incident or not.)}

The length of these flagging window time slots, $T_{FW}$, can be tuned to essentially control the level of misbehavior certainty \sysname requires before a vehicle is banned.
A long $T_{FW}$ can make sure that downvotes to a vehicle comes from multiple events, increasing the certainty that a vehicle is truly misbehaving, before it gets banned. This, however, results in a longer time to ban, and therefore higher reliance in the \textit{untrusted} state for temporary exclusion of potentially misbehaving vehicles. Recall that vehicles are able to gain and lose as much reputation in a single flagging window, while only gaining a yellow flag (if said vehicle has no prior flags). While exceedingly impractical, a crafty attacker may be able to use this time period, as well as a number of malicious vehicles with available upvotes, to stage a small, extremely targeted attack, as the yellow flag does not, on its own, prevent vehicles from participating in the network. \revise{(We discuss more about such \textit{upvote collusion attacks} in \secref{sec:preventing-collusions}.)}

However, if $T_{FW}$ is set too low, \sysname can mistakenly split downvotes from the same event into multiple ones, resulting in potential nuisance bans and easier denial-of-service attacks. We discuss our empirical derivation of $T_{FW}$ more in \apdxref{appendix:param-selection}.

When a flag is given, it stays on the vehicle's record at the SA for an amount of time we call the \textit{timeout} interval ($T_{TI}$). This is calculated by the SA on a per-vehicle basis, taking into account the vehicle's count of previous bans as a factor for multiplicatively lengthening $T_{TI}$ as the number of bans increase. The intuition is that vehicles that have been previously banned are considered less trustworthy, and therefore on subsequent bans should be penalized more.
By allowing yellow flags to \textit{fall off} a vehicle, we increase the robustness of \sysname by ensuring that rare edge cases that causes perception algorithms in vehicles to register an invalid object does not result in an immediate ban of vehicles if it happens only once. We illustrate this more in \figref{fig:cavsec-voting-state_machine}.

\parab{Banning. }
Once a red flag has been given for a vehicle, it becomes banned. The vehicle's reputation is reset to the lowest value, and the vehicle is then excluded from the network. In this state, the SA will revoke \circled{8} all of the vehicle's V2X certificates and refuse to issue more until the vehicle's red flag expires, preventing a malicious actor from performing further attacks.

Note that only upon a ban will \sysname increase the ban counter, and therefore, the timeout $T_{TI}$ for a specific vehicle. This is to ensure that punitive lengthening of the timeout is only applied when we are certain that the vehicle is misbehaving, and not upon false votes or small blips in the system.

The vehicle will then get a notification from the SA notifying it of the ban. It can then notify the owner, who can correct the malfunction. The vehicle may also voluntarily disable all data sharing features until the owner has remedied the issue.

\parab{Recovering from the Banned State.}
A vehicle may get restored to a fully-trusted state after a ban in two ways:

\parae{Self-certification: }
\sysname allows owners/operators of a vehicle to self-certify that they are fit to return to service, after waiting until the enforced timeout period ($T_{TI}$) has elapsed. After self-certification, the vehicle can start broadcasting data again, but will not become fully trusted in the network until it has gained enough upvotes (its reputation reaches $N_{thresh}$). This allows for the system to work independently in areas with no certification stations, or if the owner of the vehicle would prefer to work on their own vehicles. This approach does not reset the vehicle's previous ban counter, resulting in progressively longer timeout period $T_{TI}$ for subsequent bans.

\parae{Re-certification: }
Similar to a vehicle inspection, we envision that a vehicle owner can choose to bring the vehicle in to an authority-approved re-certification station, where in-depth inspections and certification testing can be performed. If the vehicle passes the re-certification process then it is immediately restored to fully-trusted status, bypassing the imposed timeout. The vehicle's previous bans counter is also reset to zero during this process, allowing the authority-approved re-certification process (which we envision to be rigorous) to restore vehicles that have previously experienced severe issues.

\parab{Storing and Distributing the Trust Metric. }
\sysname encodes the trust metric within each Vehicle Pseudonym Certificate (VPC) that a vehicle uses to sign V2X data, as an extended attribute of the certificate. Upon a change of trust status, the vehicle's old set of short-lived pseudonym certificates are revoked (if they are still valid), and new ones are sent out in place. This ensures that validation can be done relatively quickly in-situ (verifiers of information can simply validate the certificate) without \textit{calling the SA} for every single vehicle and for every single validation, \textit{which is inefficient and may not be possible in areas of limited connectivity}.

\input{figure_latex/states}

\subsection{Verifying data in-situ}
\label{sec:verifying-data}

\parab{Certificates. } For secure cooperative perception, \sysname's SA issues a set of certificates to each vehicle, which includes:

\begin{itemize}[leftmargin=*]
    \item \textbf{Vehicle Identity Certificate (VIC)} \circled{9}, a long-lived (lifetime) certificate for mutual authentication between the vehicle and SA, for voting and pseudonym requests.
    \item \textbf{Vehicle Pseudonym Certificates (VPC)} \circled{10}, a set of short-lived ($\le 24$ hour validity) certificates used by the vehicle to sign V2X data. As they are pseudonymous and can be shuffled, this helps with peer privacy. VPCs are also used as a vessel for the  SA to securely let vehicles self-announce their trust state to neighbors without infrastructure assistance.
\end{itemize}

VPCs are handled similarly to methods defined by the ETSI~\cite{ETSI_ITS_TS_103_097}, optionally with an extension such as \cite{ifal}. \sysname's centralized SA architecture mitigates Sybil attacks, as the SA always process votes and revocations with actual identities.

\revise{Once a vehicle is on the road, a VPC selection and rotation mechanism, such as \cite{mixzones,changing-certs-no-good,BSPpseudonym}, is used to ensure privacy. We note that \sysname is designed to be VPC algorithm agnostic, allowing for future privacy-enhancing functionalities to be incorporated without requiring a major design change.}

\parab{Verifying peer-provided data. } In a \sysname-protected network, whenever a vehicle receives data through a V2X message from other vehicles, \revise{a verifier \circled{11} first checks} that the message's signature is valid (\ie signed by a VPC that is not revoked). Then, it performs a freshness check to make sure that the beacon is not stale, delayed, or came from a vehicle with an inaccurate clock. Finally, it checks the corresponding VPC for the trust metrics, which allows it to separate incoming messages as coming from either trusted, or untrusted vehicles.

Data from untrusted vehicles are separated out, to be used only for verification and voting purposes. Data from trusted vehicles are forwarded to a \textit{majority view} algorithm \circled{2}, where checks are performed to ensure that received data is consistent, by comparing it with data from other peers and rejecting the minority. Note that this majority view comparison \revise{can happen} passively through local data aggregation, without requiring additional communication with neighboring vehicles.

Cooperative perception protocols (\eg \cite{autocast,emp}) already performs transformation of data from every vehicle into global coordinates. A low-overhead consistency checker (\eg \cite{CADConsistencyChecking,plausibilitycheck}) can be used to verify that objects received from multiple vehicles agree with each other to a reasonable threshold, allowing for an efficient \textit{majority view}, a key component (as we will discuss in \secref{sec:majorityview_ablation}) that contributes to the performance of \sysname. Additionally, vehicles may also incorporate additional consistency checks on other V2X messages (\eg location data), using an algorithm such as \cite{gpsconsistency}, to further increase robustness.

\subsection{Prevention of vote collusions}
\label{sec:preventing-collusions}

\revise{A \textit{downvote collusion attack} is when an adversary attempts to abuse \sysname's voting protocol, tricking \sysname into falsely banning a benign vehicle. To prevent this, \sysname requires at least two downvotes in separate voting windows. Recall that a vehicle is only able to perform one downvote per $T_{IDE}$. Thus, an adversary would require at least two vehicles to perform an attack on a target vehicle $T$. While this would, indeed, exclude $T$ from the network for an amount of time $T_{TI}$, the two \textit{malicious downvoters} themselves will no longer be eligible to downvote any other vehicle for $T_{IDE}$, which (being much longer than $T_{TI}$) ensures that such attacks are not sustainable or practical at scale, as such a malicious attacker would need to acquire or compromise an increasingly larger amount of vehicles to continually perform their attack, while their victims would be restored to the network well before the attacker can repeat the attack on them (or anyone else) again.}

\revise{Similarly, for an \textit{upvote collusion attack}, an adversary attempts to use multiple vehicles to "prop up" a single attacking vehicle's trust within $T_{FW}$, attempting to work around \sysname's quick response mechanism (recall that \sysname bans vehicles as soon as another downvote is received outside of the first flagging window, so the attacker will be excluded by then). Since vehicles can only upvote once per $T_{IVE}$, we easily see how this will become even more unsustainable than the \textit{downvote collusion attack}, requiring an attacking vehicle fleet that is larger than the amount of benign vehicles in the same area just to perform an attack in a single $T_{FW}$ window.}

\subsection{Other details}
\label{sec:details}

\parab{Vehicle provisioning.}
When a vehicle is manufactured and the autonomy system is started for the first time, the vehicle's hardware security module (HSM) will securely generate the VIC key pair. The private key will never leave the vehicle's HSM, while the public key can then be exported and signed-off with the manufacturer's credentials to be signed by the SA.
This process binds the manufacturer to a vehicle identity at the SA, preventing a Sybil attack \circled{12}.
Once the SA signs the VIC, the vehicle is ready to get its own VPCs directly from the SA.

\parab{V2X communications.}
\sysname uses two in-band communication channels for direct vehicle-to-vehicle communications: a {\textit{beaconing}} channel, and a separate \textit{data} channel.
From a vehicle's perspective, at any point in time, it will pick a single pseudonym certificate from its pool of available certificates \circled{13} to be used in V2X communications, \revise{using protocols as discussed earlier in \secref{sec:verifying-data} to provide anti-tracking protections.}

\parab{Beaconing.}
With a pseudonym selected, the vehicle can then select the corresponding VPC certificate/key pair to use. Then, it broadcasts a \textit{beacon} over the beaconing channel once every \textit{broadcast interval} $T_{BBI}$. This beacon contains the VPC certificate (which contains the pseudonym UUID and trust state), a timestamp, and optionally any additional information for V2X synchronization. It is signed by the VPC private key, which prevents the beacon from being tampered in transit.

\sysname beacons allows surrounding vehicles to \textit{learn} the pseudo-identity of a soon-to-be peer. Considering that vehicles move around and are only in contact with each other for a short period of time, it is space-impractical to pre-load all possible pseudonym certificates into every participant vehicle.

Similarly, including the certificate as part of every V2X message is not ideal, as certificates, even with elliptic-curve (EC) cryptography, can be large (up to 709 bytes). Including certificates into every frame wastes precious airtime and frame space, as V2X standards such as WAVE \cite{ieeewave} utilizes the Ethernet frame standard, which maxes out at 1500 bytes.

Introducing session setup for larger-size data transport is non-ideal in such a dynamic network, as it can introduce the $n$-way handshake problem, causing significant overhead. Hence, we opted to design \sysname such that it uses broadcast beacons to inform neighboring vehicles about its current identity. Every vehicle that receives this beacon (which contains the pseudonym ID as well as the certificate) can cache it to be recalled later to validate data frames that a vehicle may send out.

The optimal value for $T_{BBI}$ is a balance between minimizing the time a vehicle need to recognize and \textit{trust} other vehicles and minimizing bandwidth requirements. We discuss our derivation of $T_{BBI}$ by statistical traffic data in \apdxref{appendix:param-selection}.

\parab{Data transmission.}
When a vehicle would like to share any data with those around it, it will generate a spontaneous message. This message is then timestamped, and signed \circled{14}, before it is sent out through the V2X radio. As EC signatures are small (up to 132 bytes with P-521, smaller on others), and the only other piece of data needed in the header is the pseudonym ID (16 bytes), we keep the overhead small.

\parab{Glossary of parameters.} 
\tabref{tab:online-learnable-parameters} and \tabref{tab:ground-truth-parameters} summarizes the tunable parameters introduced earlier in our design, as parameters that can be adjusted based on \textit{in-situ} measurements, and those that are configured based on analysis of data where ground truth information is made available (\eg from simulations or statistical datasets), respectively.

Note again that \sysname's design attempts to make reasonably trustable assumptions from multiple, potentially untrusted data points, while at the same time attempting to limit direct collection of sensitive data (\eg precise location information), leading to a larger number of parameters in general. In the interest of space, we describe the selection process, as well as our derivation for these parameters in \apdxref{appendix:param-selection}.

\input{tables/parameter-glossary}

%% file: figure_latex/system.tex
\begin{figure}
    \includegraphics[width=0.95\columnwidth]{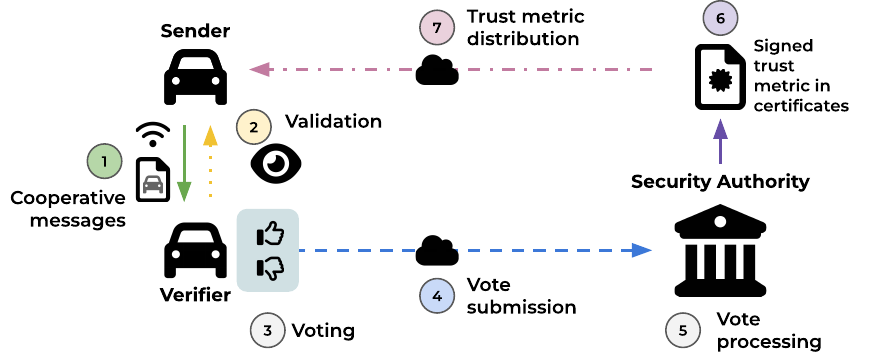}
    \caption{\revisetwo{\sysname system workflow.}}
    \label{fig:sysdesign-vehicle}
\end{figure}

%% file: figure_latex/arch.tex
\begin{figure*}[t]
    \centering
    \includegraphics[width=0.95\textwidth]{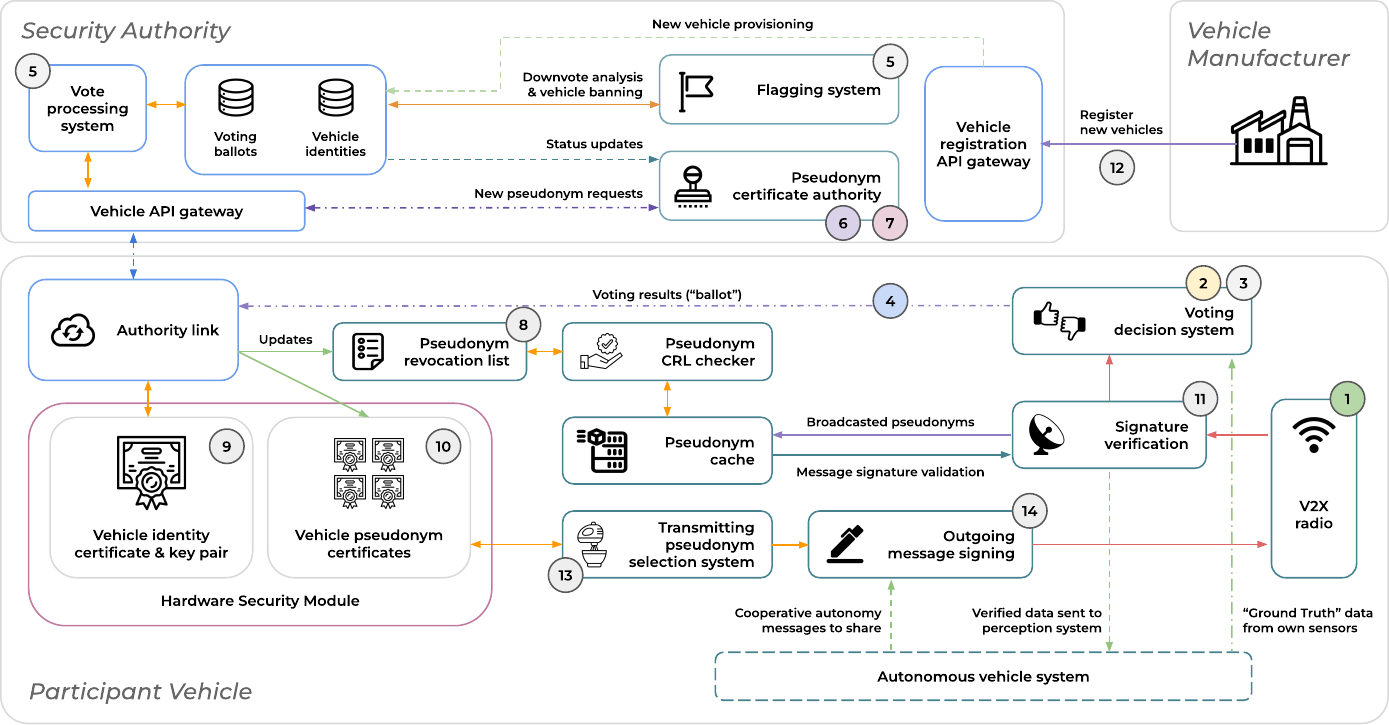}
    \revise{\caption{\revisetwo{\sysname architecture at a component level, showing how individual pieces fits together.}}}
    \label{fig:cavsec-system-components}
\end{figure*}

%% file: algorithms/sa-vote-processing.tex
\begin{algorithm}
    \caption{{SA vote processing}}\label{alg:sa-vote-acceptance}
    \begin{algorithmic}[1]

        \floatingensure{$V_{vote}$ and $V_{target}$ are actual identities}
        \floatingensure{$V_{vote} \neq V_{target}$ and $V_{vote}$ is trusted}

        \State $B \gets \text{vote ballot from } V_{vote}$

        \State $T_{now} \gets $ current time at the authority
        \floatingensure{$T_{now}$ - timestamp($B$) $\le T_{vote}$}

        \If{$B$ is upvote}

            \floatingensure{$(V_{vote}, V_{target}) \notin UL_{[T_{now}, T_{now} - T_{IVE}]}$}

        \ElsIf{$B$ is downvote}
            \floatingensure{$(V_{vote}, V_{target}) \notin DL_{[T_{now}, T_{now} - T_{IVE}]}$}
            
            \floatingensure{$(V_{vote}, *) \notin DL_{[T_{now}, T_{now} - T_{IDE(V)}]}$}
        \EndIf

        \If{$B$ is upvote}
            \State add $(V_{vote}, V_{target})$ to $UL$
            \State increase reputation for $V_{target}$
        \ElsIf{$B$ is downvote}
            \State add $(V_{vote}, V_{target})$ to $DL$
            \State decrease reputation for $V_{target}$
        \EndIf

        \For{all non-banned $V_{target}$ added to $UL$ or $DL$}
            \If{score($V_{target}$) $< N_{thresh}$}
                \State $V_{target}$ is \textit{untrusted}
            \EndIf
        \EndFor

    \end{algorithmic}
\end{algorithm}

%% file: figure_latex/states.tex
\begin{figure*}[t]
    \centering
    \includegraphics[width=16.5cm]{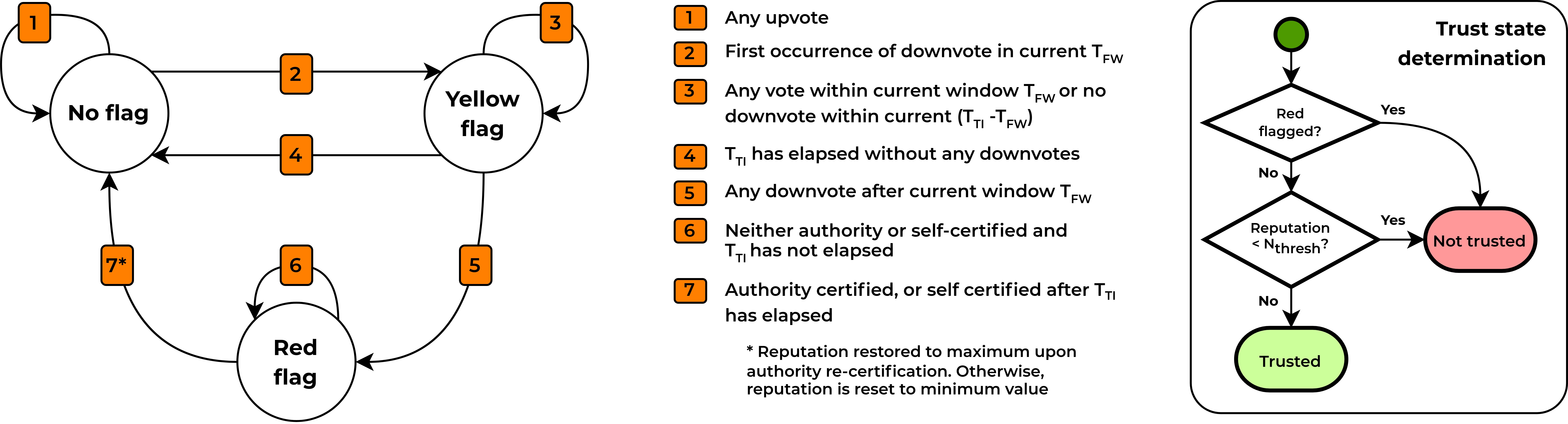}
    \caption{State diagram for \sysname (transition descriptions in middle). Trust state determination (right) is run after every transition.}
    \label{fig:cavsec-voting-state_machine}
\end{figure*}

%% file: tables/parameter-glossary.tex
\begin{table}
    \resizebox{\columnwidth}{!}{%
        \begin{tabular}{@{}lll@{}}
            
        \toprule
        \textbf{Parameter}      & \textbf{Name}             & \textbf{Description}                                                                                              \\ \midrule
        $T_{IVE}$                & Inter-vote epoch                & \begin{tabular}[c]{@{}l@{}}A vehicle $V_i$ can vote for \\ vehicle $V_j$ once per $T_{IVE}$ \end{tabular}                      \\
        $T_{IDE}$                & Inter-downvote epoch            & \begin{tabular}[c]{@{}l@{}}A vehicle $V_i$ can downvote any \\ vehicle once per $T_{IDE}$ \end{tabular}                     \\
        $T_{vote}$              & Vote submission limit     & \begin{tabular}[c]{@{}l@{}}Maximum time between beacon \\ generation and vote submission\end{tabular}             \\
        
        $T_{TI}$               & Timeout interval      & \begin{tabular}[c]{@{}l@{}}Multiplicatively increasing timeout \\ interval for misbehaving vehicles \end{tabular}          \\ \bottomrule
        \end{tabular}%
        }
        \vspace{1mm}
        \caption{Glossary of online-learnable parameters.}
        \label{tab:online-learnable-parameters}
\end{table}

\begin{table}
    \resizebox{\columnwidth}{!}{%
        \begin{tabular}{@{}lll@{}}
            
        \toprule
        \textbf{Parameter}      & \textbf{Name}             & \textbf{Description}                                                                                              \\ \midrule
        $T_{FW}$                & Flagging window duration  & \begin{tabular}[c]{@{}l@{}}Duration for event-grouping\\ in the flagging algorithm\end{tabular}                   \\
        $T_{BBI}$               & Beacon broadcast interval     & \begin{tabular}[c]{@{}l@{}}Security beacon / certificate \\ broadcasting interval\end{tabular}                    \\ 
        $N_{thresh}$            & SA reputation threshold        & \begin{tabular}[c]{@{}l@{}}"Trusted" threshold for \\ SA-internal reputation score\end{tabular}                        \\ \bottomrule
        \end{tabular}%
        }
        \vspace{1mm}
        \caption{Glossary of parameters based on ground truth analysis.}
        \label{tab:ground-truth-parameters}
\end{table}

%% file: sections/experimental-evaluation.tex
\section{Experimental Evaluation}
\label{sec:experimental-evaluation}

In order to stress-test the operation of our system at scale, we created a simulation environment that runs on top of SUMO \cite{sumo2018}, a simulation tool capable of realistic, city-scale road network simulation, utilizing two large-scale scenarios \revise{with a time slice resolution of 1 second}:

\begin{itemize}
    \item Berlin SUMO traffic dataset \cite{BerlinSUMO}, {a manually fine-tuned dataset based on \cite{berlin-best-scenario}, representing traffic in the city of Berlin} (with a map size of $800 \text{ km}^2$) 
    \item ETH Zurich Boston dataset \cite{BostonDataset}, {a simulation of urban traffic in the city of Boston based on real-world statistics}
\end{itemize}

Misbehavior probabilities are chosen to exceed real-world expectations (as we later derive, based on real empirical data, in \secref{sec:risk-analysis}) to cover even highly unlikely scenarios, such as reduced enforcement against owners of vehicles with bad sensors, or sophisticated malicious attacks involving a large number of colluding attackers. We thus run the simulation with 0.2\%, 2\%, 5\%, and 20\% of misbehaving vehicles, ensuring resilience of \sysname under a range of conditions, from typical to extreme.

\revise{In our experiments, we assume a simplified model of broadcast communications where vehicles will be able to communicate if they are within range of each other, chosen to be 400m based on \cite{QualcommCV2X}. Propagation delays and packet loss (or channel congestion) are not modeled in order to focus on the application rather than the lower network-stack layers, however, based on analytical results on V2X performance \cite{cv2x_mode4_performance,5gnr_performance} and our 1-second sampling rate, we assume that any retransmissions, if required, will be completed within the sample time slot.}

Then, we perform ablation studies to determine the contribution of our 3-state trust design, the \textit{majority view} algorithm, \revise{as well as per-vehicle $T_{IDE}$ scaling} to the performance of \sysname.

\revise{Finally, we perform evaluations on crytographic performance of \sysname, and present an example \textit{vehicle state timeline} to help visualize how vehicle state transitions work under \sysname.}

\subsection{Large scale experiment}
\label{sec:large-scale-experiment}

\input{figure_latex/comparative-analysis}

\parab{Methodology. }
We conducted a large scale experiment on \sysname, along with systems modeled after \VR (a pure reputation based system) \cite{VanetReputation} and \TruPercept (a short-window, continuous majority weighting system) \cite{TruPercept}, using both scenarios over a simulated rush hour over 700 seconds. This duration was chosen to be computationally viable, while adequately capturing systems' reactions. At any moment, there is an average of over 27,000 vehicles in the simulation, \revise{all with V2X capability.} Vehicle sensor ranges are conservatively set to 30m, based on data from state-of-the-art sensors \cite{OusterLidar,TeslaAutopilot} and real-world intersection sizes.

\input{tables/experiment-parameters}

For the purpose of our experiments, the online learnable parameters of \sysname (\tabref{tab:online-learnable-parameters}) are computed. More details of said computation, as well as parameters whose value is based on ground truth analysis (\tabref{tab:ground-truth-parameters}), are explained in detail in \apdxref{appendix:param-selection}. \tabref{tab:experiment-parameters} shows the values used in the experiments.

We initialize the simulation with a populated map, with all vehicles behaving as trustworthy peers in the network. Then, for subsequently arriving vehicles, we randomly and uniformly turn some into misbehaving peers - either as a \textit{bad sensor vehicle}, or a \textit{flip-flopping vehicle}. 

Bad sensor vehicles are those that have consistently inaccurate perception of the environment, due to a faulty sensor. As such, they will broadcast bad data, as well as downvote all other vehicles due to their misunderstanding of the environment.

Flip-flopping vehicles, on the other hand, are assumed to be malicious actors that switch between sending good and bad data to try and evade detection. They would perform voting truthfully otherwise for stealthiness (or to keep the one downvote they have for an attempt at a malicious downvote later).

For every misbehaving vehicle, we also assigned a time-delay, uniformly distributed against the half-life of the vehicle (i.e., based on when it entered and would exit the simulation), to distribute misbehaving vehicles spatially in the map.

For each ratio of misbehaving vehicles (i.e., those that will send \textit{bad} data one way or another), we collected data from multiple independent runs of the simulations on both cities.

\begin{figure}[t]
    \centering
    \includegraphics[width=0.85\columnwidth]{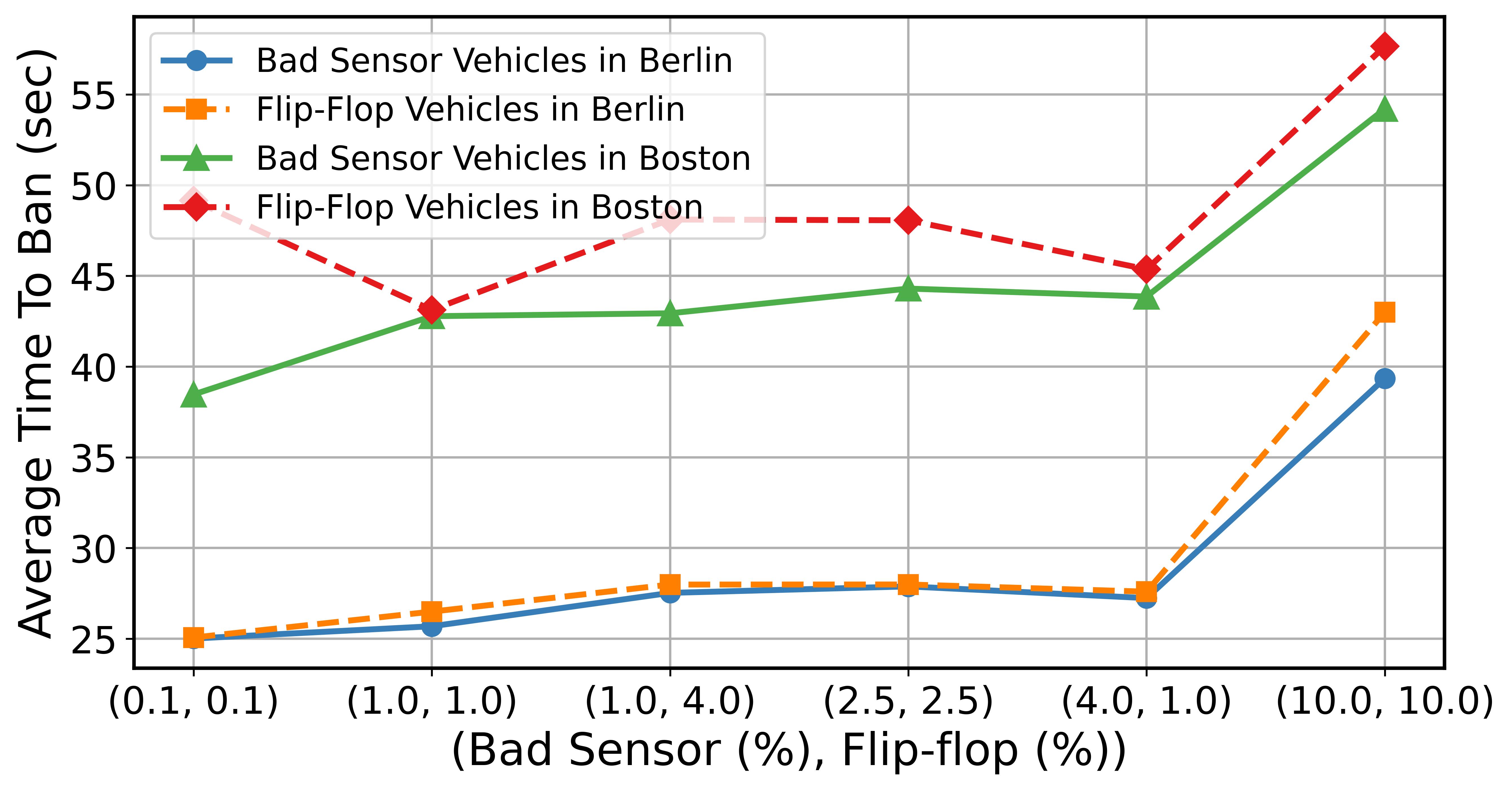}
    \caption{Average misbehaving vehicle ban times of \sysname}
    
    \label{fig:average-time-to-ban-berlin-boston}
\end{figure}

\parab{Results. }
We provide our experimental results on:

\begin{itemize}[leftmargin=*]
    \item Percentages of bad messages that are let through (filtering failure; false negative \%) (\figref{fig:fn-berlin} and \figref{fig:fn-boston})
    \item Percentages of good messages that are dropped (overly sensitive filter; false positive \%) (\figref{fig:fp-berlin} and \figref{fig:fp-boston})
    \item Absolute time required for \sysname {(as the only system with banning)} to ban a misbehaving vehicle (\figref{fig:average-time-to-ban-berlin-boston})
\end{itemize}

Our findings in \figref{fig:all_plots} reveal that \sysname excels at filtering bad messages from misbehaving vehicles by an average of 230x, when compared to the next best system (\VR, a pure reputation-based approach), greatly reducing the probability that a bad message will affect control decisions.

For an example, take the case where we have $2.5\%$ of vehicles with a bad sensor, and $2.5\%$ as an attacker, for a total misbehavior rate of $5\%$. We observe that \sysname significantly reduces the ratio of false negatives (257x) compared to the false-negative tuned ($\psi_{nf} = 2$) version of \VR, while at the same time only making a small trade-off in false positives (3x), compared to the false-positive tuned ($\psi_{nf} = 100$) version of \VR, showing that \sysname, as presented, can provide the best balanced protection as measured by both metrics. We believe that for autonomous vehicles, where safety is paramount, accepting this tradeoff is a reasonable approach.

Further, we observed in \figref{fig:average-time-to-ban-berlin-boston} that \sysname's banning feature responds quickly. On average, a misbehaving vehicle is excluded from the network in less than 58 seconds, which helps prevent the spread of bad messages. Although in certain edge cases a vehicle may take a long time to become banned, it's important to note that such a delayed ban does not necessarily lead to adverse outcomes - as \sysname also includes other defensive mechanisms, from \textit{majority view} to the \textit{untrusted} state that provides fast-acting responses to misbehavior even when a vehicle has not been banned yet.

The edge \sysname has over \VR ~{(in terms of lower false negatives and false positives as shown earlier)} is largely due to the utilization of a \textit{majority view} mechanism, which enables vehicles to passively verify the accuracy of information received from others when there is multiple sources of information available. {From observation of real-world datasets, we found that it is common for multiple vehicles to simultaneously detect the same object on the road. By exploiting this whenever possible, \sysname can discard bad messages in cases where purely reputation-based systems cannot.}

We also observed that long-term memory (and banning) can also help increase robustness. While \TruPercept also employs majority view, it suffers from a 'short-term memory' issue, as unlike \sysname, it does not maintain a long-term record of a vehicle's trustworthiness, leading to a quick reset of its memory of a vehicle's past misbehavior in just a few minutes.

Additionally, we benchmarked our in-situ pseudonym certificate verification scheme and found that this can be done in less than 10 milliseconds on modern processors, highlighting \sysname's applicability for real-world, real-time deployments.

\subsection{Ablation study on 3-state trust design and trust threshold}
\label{sec:3states_ablation}

{We conducted an ablation study using the Boston dataset, where we compare \sysname (with 3-state trust) with a 2-state trust variant (which only has the \textit{trusted} and \textit{banned} states).}

Our results in \tabref{tab:res-ablation-3states} shows that a 3-state design significantly reduces the percentage of false negatives, compared to a 2-state design (by up to 13x), while showing no significant changes in false positives, indicating that a 3-state design contributes positively to the performance of \sysname (by enabling faster responses without affecting false ban prevention).

\subsection{Ablation \revise{studies} on majority view}
\label{sec:majorityview_ablation}

\input{tables/3state-ablation.tex}
\input{tables/majority-view-ablation.tex}
\input{tables/majority-view-purevote.tex}

To determine the effectiveness and benefits of integrating the \textit{majority view} algorithm, we performed multiple shorter experiment runs under the Boston scenario to compare regular \sysname with a modified version without majority view.

Our results in \tabref{tab:res-ablation-mv} shows that with regular \sysname, we observe a significant (up to 2.6x) reduction in false negatives, over a variant with no \textit{majority view}.

\revise{We also performed the inverse experiment, by removing \sysname's reputation mechanism and instead relying entirely on \textit{majority view} (in essence comparing \sysname to the ideal pure voting-based method). Our results in \tabref{tab:res-ablation-purevote} shows that a combined voting and reputation design outperforms a pure voting-based system in ensuring bad messages are dropped.}

\revise{
\subsection{Ablation study on $T_{IDE}$ scaling and sensitivity analysis}
}
\label{sec:tide_scaling}

\begin{figure}[]
    \centering
    \includegraphics[width=0.95\columnwidth]{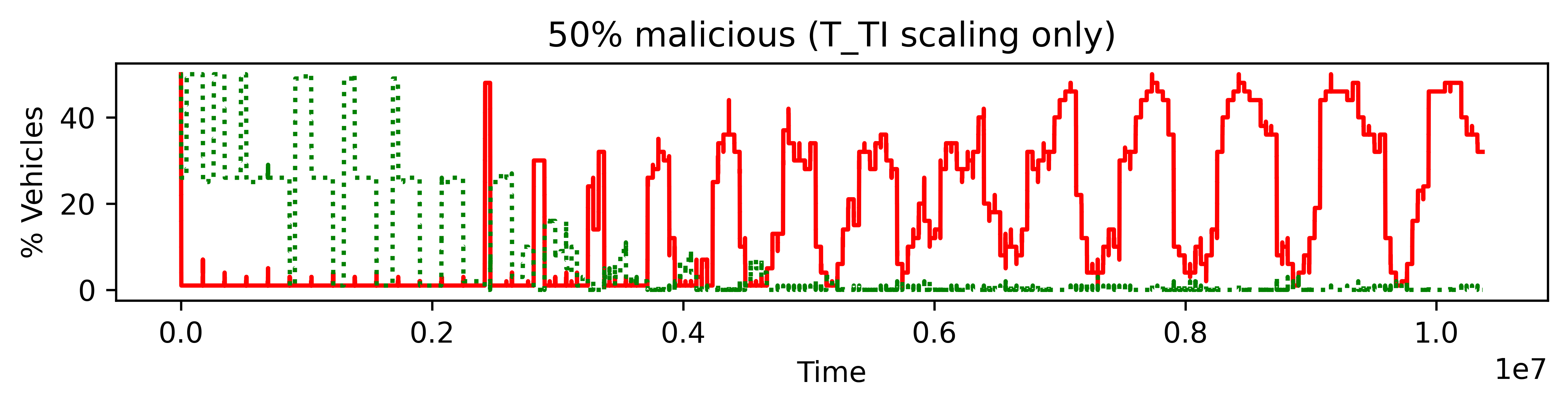}
    \includegraphics[width=0.95\columnwidth]{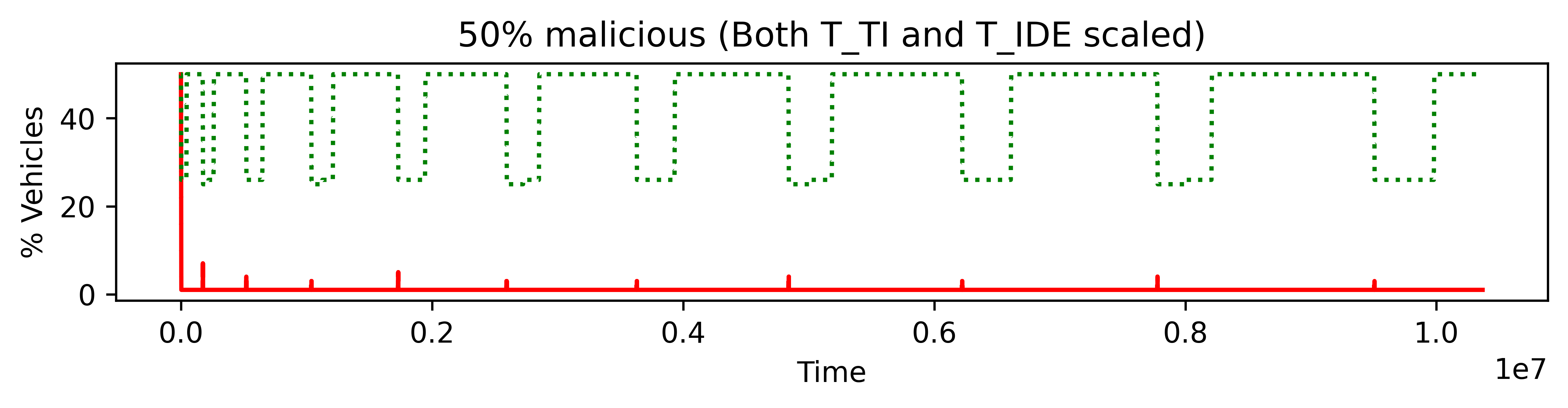}
    \caption{Comparison result excerpt for a \sysname design that only scales $T_{TI}$, vs. a design that scales both $T_{TI}$ and $T_{IDE}$.}
    \label{fig:tide_tti_scaling}
\end{figure}

\revise{To determine the efffectiveness and benefits of scaling $T_{IDE}$ on a \textit{per-voter} basis as a denial-of-service prevention mechanism, we perform a voting-only experiment on a longer time scale, within a closed environment of finite vehicles. We then vary the ratio of \textit{malicious vehicles} (who all collude with each other with perfect timing to try and ban as many benign vehicles as possible) to \textit{benign} vehicles, and compare a design that scales both $T_{TI}$ and $T_{IDE}$ to another that only scales $T_{TI}$. \figref{fig:tide_tti_scaling} shows an excerpt of how the vehicle population evolves over time (dotted line and solid line represent the percentage of \textit{benign} vehicles that are active (\ie not banned) and the percentage of \textit{malicious} vehicles with a downvote available (\ie ready to attack) respectively), showing that by scaling both $T_{IDE}$ and $T_{TI}$, \sysname mitigates long-term denial of service attack by ensuring vehicles will return to service faster than they can be falsely banned by malicious actors. We note that near-identical results are observed with a varying ratio of malicious vehicles from $5\%$ to $50\%$, showing insensitivity to the ratio of malicious to benign vehicles.}

\revise{
\subsection{Cryptography evaluation}
}
\label{sec:certeval}

\revise{
\parab{Data verification time. }
We validated \sysname's certificate, signature, and trust state verification time by implementing these steps as a standalone single-threaded test application, and testing them on 3 systems with different processors that are likely representative of different classes of processors that may be found in different levels~\cite{J3016} of autonomous vehicles:
}

\revise{
\begin{itemize}[leftmargin=*]
    \item Intel Core i7-11800H (2.3 GHz; laptop), representative of a lower-end vehicle class, e.g., those at SAE level 3.
    \item AMD Threadripper 3970X (3.7 GHz; desktop), representative of server-class processors reportedly deployed in SAE level 4 autonomous vehicles \cite{WaymoXeon}.
    \item AMD Ryzen 9 7900X (4.7 GHz; desktop), representative of future SAE level 4 or 5 autonomous vehicles that will run processors with higher base clocks.
\end{itemize}
}

\begin{figure}[t]
    \centering
    \includegraphics[width=7.5cm]{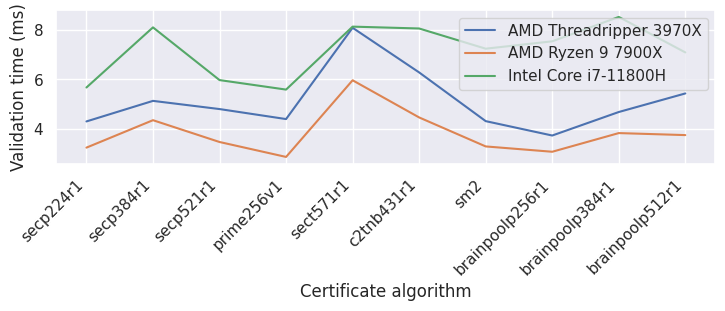}
    \caption{\sysname validation time evaluation.}
    \label{fig:cert-validation-time}
\end{figure}

\revise{
Figure \ref{fig:cert-validation-time} summarizes our findings. We observed $<10 ms$ verification time on all processors and algorithms, showing that \sysname can be integrated into autonomous driving stacks without significantly increasing end-to-end latency. 
}

\revise{
\parab{Certificate sizes. }
We also performed size analysis of \sysname certificates (carrying the signed UUID and trust attribute). Table \ref{tab:certsizes} summarizes our findings, showing that reasonably secure EC certificates are small enough to be transmitted in one packet, making beaconing of certificates possible.
}

\begin{table}
  \centering
  \begin{tabular}{@{}lll@{}}
      \toprule
      Algorithm       & \begin{tabular}[c]{@{}l@{}}Security strength (bits)\end{tabular} & \begin{tabular}[c]{@{}l@{}}Certificate size (bytes)\end{tabular} \\ \midrule
      secp224r1       & 112                       & 501                       \\
      brainpoolP256r1 & 128                       & 530                       \\
      secp384r1       & 192                       & 611                       \\
      brainpoolP512r1 & 256                       & 709                       \\ \bottomrule
    \end{tabular}
  \vspace{1mm}
  \caption{\sysname certificate sizes with different algorithms.}
  \label{tab:certsizes}
\end{table}

\revise{
\subsection{Vehicle state transitions}
}
\label{sec:vehicle_state_transitions}

\begin{figure}
    \vspace{3mm}
    \centering
    \includegraphics[width=0.65\columnwidth]{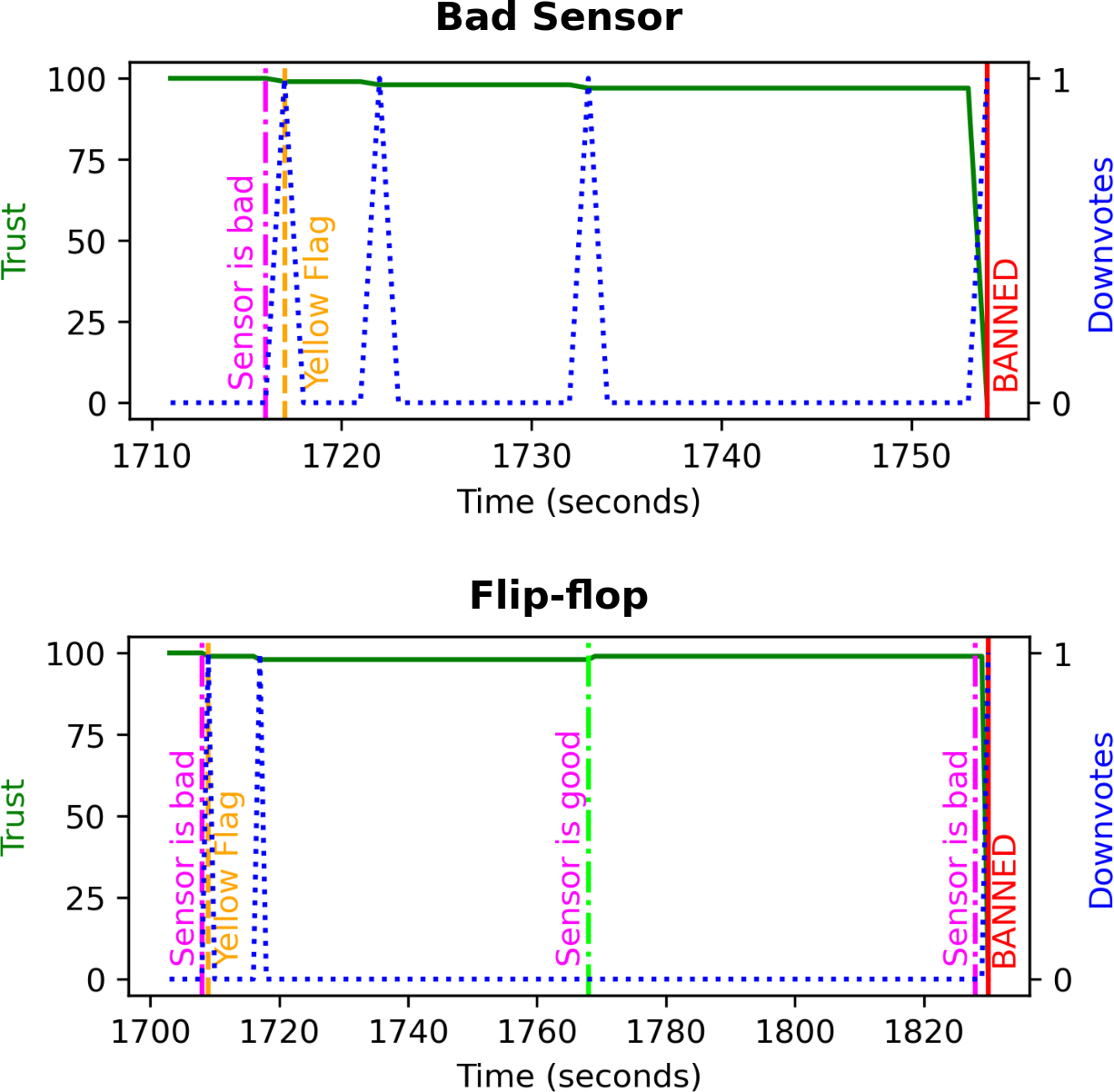}
    \caption{Vehicle state timeline example cases}
    \label{fig:svt-example-badsensor}
\end{figure}

\revise{To help visualize vehicle state transitions, we extract an example \textit{state timeline} from two example vehicles in an instance of our large-scale experiment (\secref{sec:large-scale-experiment}) and present them in \figref{fig:svt-example-badsensor}. Here, we show important events (\eg the first moment that a sensor went bad or an attack has started, and flag state change events) as vertical lines, alongside plots for the vehicles' trust score as well as the number of received downvotes.}

\revise{In the bad sensor case, we first observe that a sensor on vehicle $A$ goes bad at $T = 1716$, which is first detected by another vehicle at $T = 1718$. Immediately, this reduces $A$'s reputation and a yellow flag is issued. In the next $T_{FW}$ window (in this case, 20 seconds), two downvotes are received and $A$'s reputation is reduced accordingly, however, the vehicle remains in the yellow flag state until the next window. Eventually, outside of that window, $A$ gets downvoted again at $T = 1755$, resulting in a red flag and thus a ban from the network.}

\revise{A similar scenario occurs for the flip-flop vehicle $B$, with the difference being that the vehicle regains some reputation after it stops the attack and starts sending accurate data again (at $T = 1768$). However, once $B$ starts an attack on a new victim (at $T = 1830$), it gets banned as soon as the next attack is detected as $B$ already has a yellow flag from its earlier attack.}

%% file: figure_latex/comparative-analysis.tex
 \begin{figure*}[!t]
   \centering
   \begin{subfigure}[b]{0.43\textwidth}
     \includegraphics[width=\textwidth]{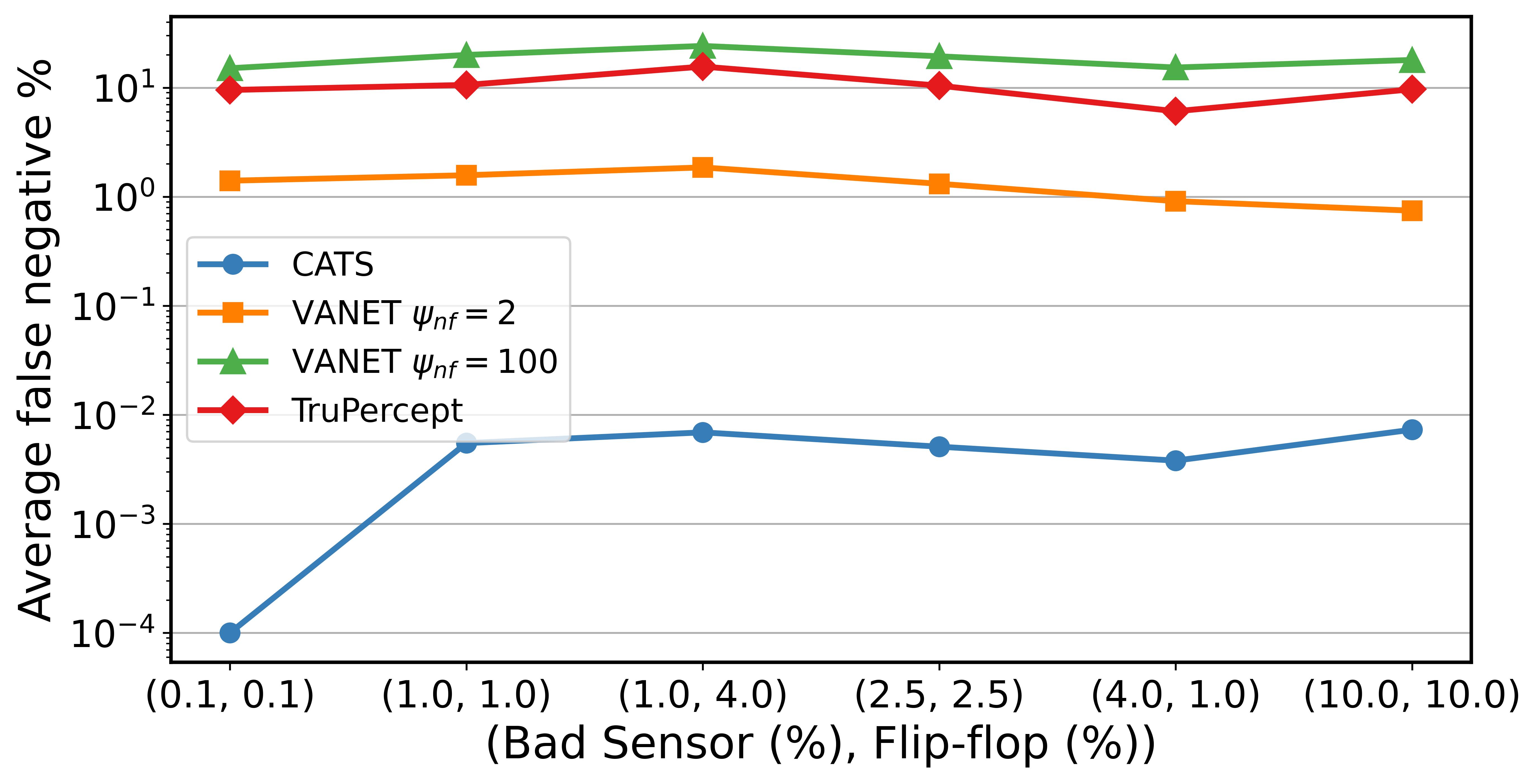}
     \caption{Average false negative message \% in Berlin (log-scale).}
     \label{fig:fn-berlin}
   \end{subfigure}
   \hspace{12mm}
   \begin{subfigure}[b]{0.43\textwidth}
     \includegraphics[width=\textwidth]{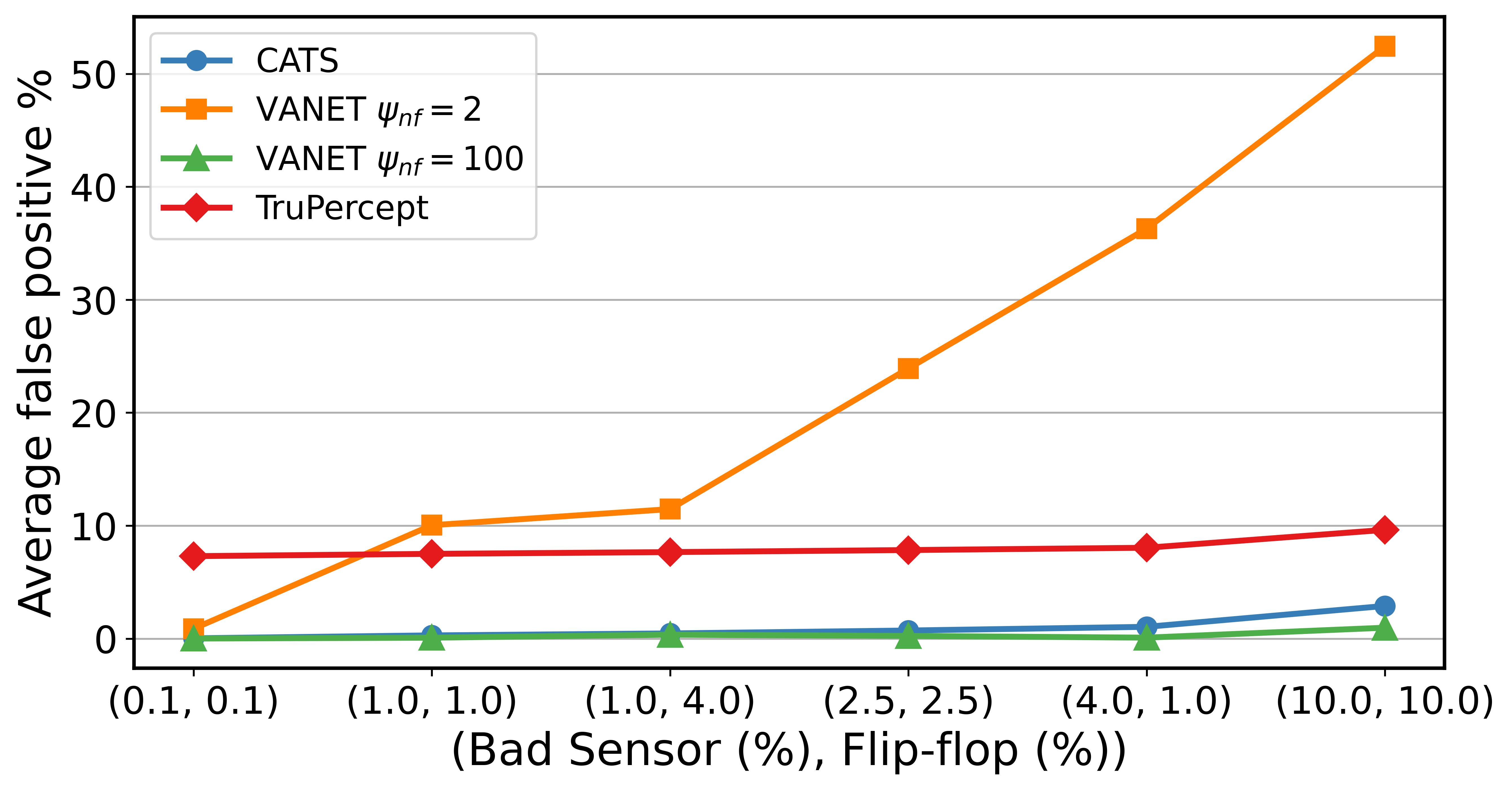}
     \caption{Average false positive message \% in Berlin.}
     \label{fig:fp-berlin}
   \end{subfigure}
   \begin{subfigure}[b]{0.43\textwidth}
     \includegraphics[width=\textwidth]{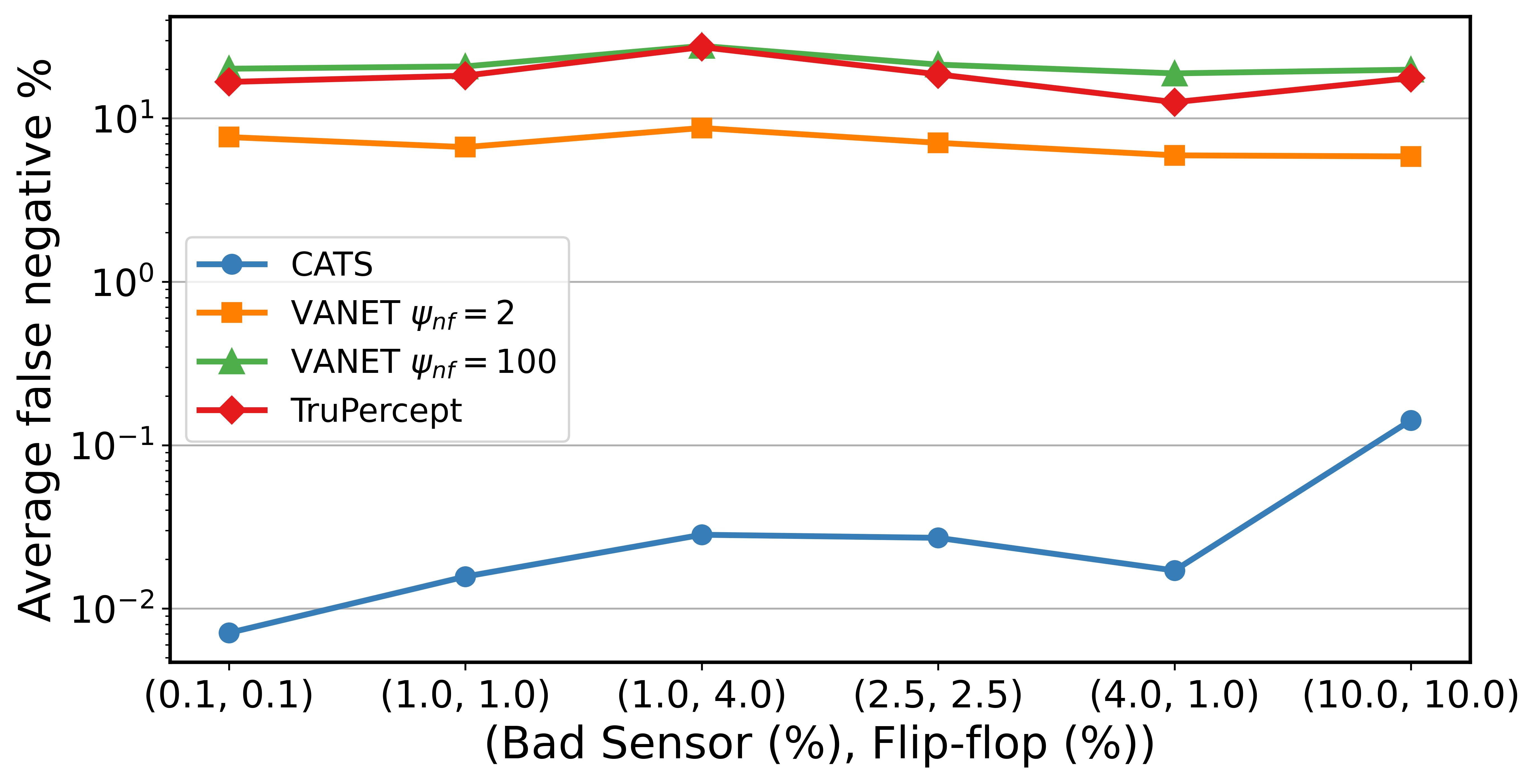}
     \caption{Average false negative message \% in Boston (log-scale).}
     \label{fig:fn-boston}
  \end{subfigure}
   \hspace{12mm}
   \begin{subfigure}[b]{0.43\textwidth}
     \includegraphics[width=\textwidth]{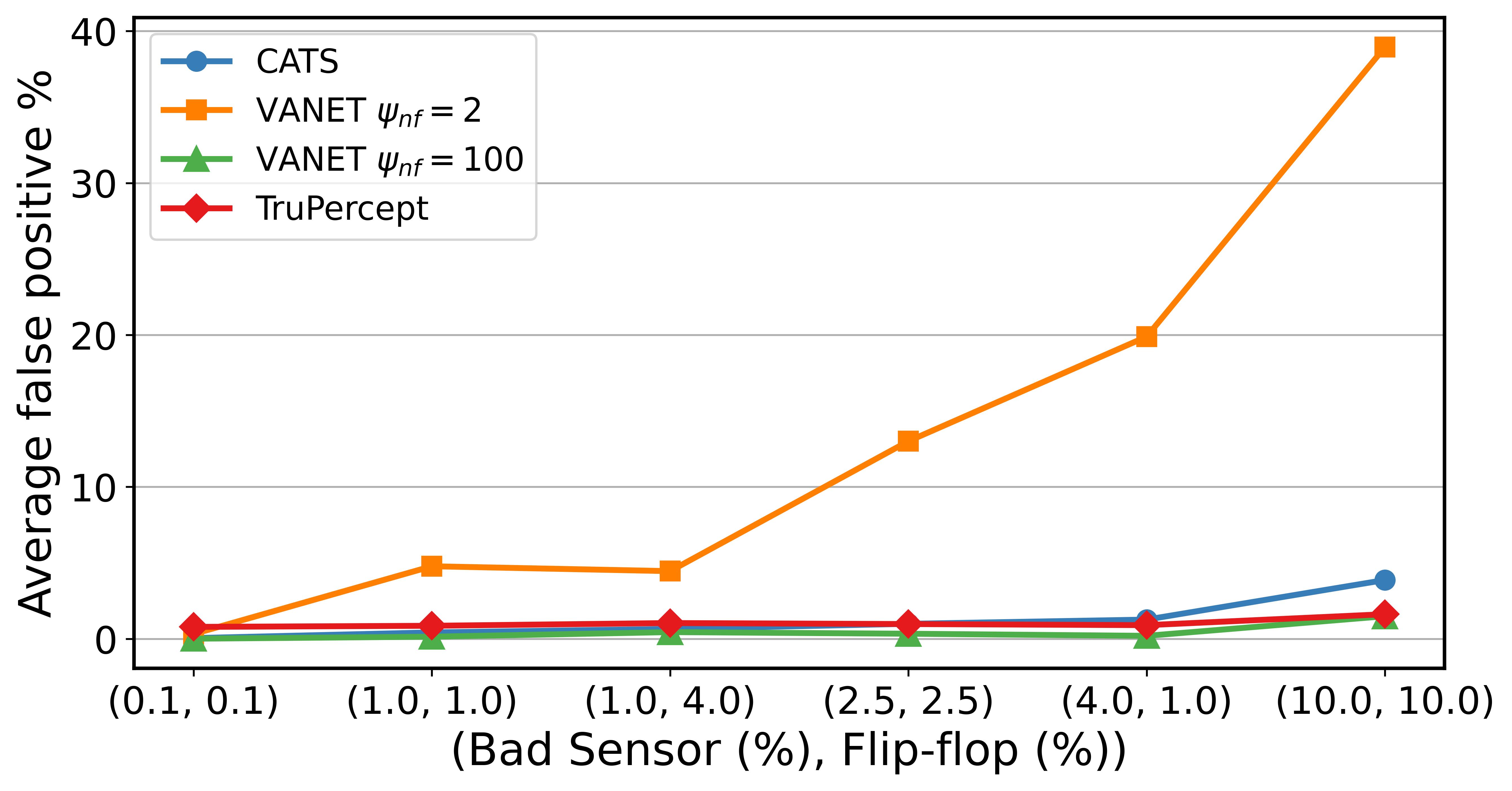}
     \caption{Average false positive message \% in Boston.}
    \label{fig:fp-boston}
   \end{subfigure}
   \caption{Comparative analysis of false negative and false positive message percentages for misbehavior detection methods (\sysname, \VR with $\psi_{nf} = 2$ and $\psi_{nf} = 100$, and \TruPercept) in Berlin and Boston traffic simulation runs.}
   \label{fig:all_plots}
 \end{figure*}

%% file: tables/experiment-parameters.tex
\begin{table}[h]
    \resizebox{\columnwidth}{!}{%
        \begin{tabular}{@{}lll@{}}
            
        \toprule
        \textbf{Parameter}      & \textbf{Name}                     & \textbf{Value used}                                                                                              \\ \midrule
        $T_{FW}$                & Flagging window duration          & 20 seconds                    \\
        $T_{BBI}$               & Beacon broadcast interval         & 1 Hz                          \\
        $N_{thresh}$            & Normalized SA trust threshold     & $0.998$     \\ \bottomrule        
        \end{tabular}
        }
        \vspace{1mm}
        \caption{Selected ground-truth parameters for experiments.}
        \label{tab:experiment-parameters}
\end{table}

%% file: tables/3state-ablation.tex
\begin{table}
        \centering
        \begin{tabular}{@{}ll|ll|ll@{}}
\toprule
\multirow{2}{*}{\shortstack{\% Bad\\sensor}} & \multirow{2}{*}{\shortstack{\% Flip-\\flop}} & \multicolumn{2}{l|}{\% False negative messages} & \multicolumn{2}{l}{\% False positive messages} \\ \cmidrule(l){3-6} 
 &  & \multicolumn{1}{l|}{2-state trust} & 3-state trust & \multicolumn{1}{l|}{2-state trust} & 3-state trust \\ \midrule
1 & 1 & \multicolumn{1}{l|}{0.5712} & \textbf{0.0430} & \multicolumn{1}{l|}{0.2900} & 0.2902 \\
1 & 4 & \multicolumn{1}{l|}{0.7589} & \textbf{0.0748} & \multicolumn{1}{l|}{0.4888} & 0.4899 \\
2.5 & 2.5 & \multicolumn{1}{l|}{0.6018} & \textbf{0.0547} & \multicolumn{1}{l|}{0.6784} & 0.6787 \\
4 & 1 & \multicolumn{1}{l|}{0.6809} & \textbf{0.0524} & \multicolumn{1}{l|}{1.0303} & 1.0303 \\
10 & 10 & \multicolumn{1}{l|}{0.8162} & \textbf{0.2660} & \multicolumn{1}{l|}{2.8106} & 2.8117 \\ \bottomrule
\end{tabular}
\vspace{1mm}
    \caption{Ablation study results for false negative and positive message percentages in a 2 vs. 3 state trust system design.}
    \label{tab:res-ablation-3states}
\end{table}

%% file: tables/majority-view-ablation.tex
\begin{table}
    \centering
    \begin{tabular}{@{}ll|ll@{}}
    \toprule
    \% Bad sensor & \% Flip-flop & With majority view & Without majority view \\ \midrule
    1             & 1            & \textbf{0.0152}             & 0.0303                \\
    1             & 4            & \textbf{0.0289}             & 0.0770                \\
    2.5           & 2.5          & \textbf{0.0369}             & 0.0679                \\
    4             & 1            & \textbf{0.0360}             & 0.0774                \\
    10            & 10           & \textbf{0.0784}             & 0.2060                \\ \bottomrule
    \end{tabular}
    \vspace{1mm}
    \caption{False negative \% for majority view ablation study.}
    \label{tab:res-ablation-mv}
\end{table}

%% file: tables/majority-view-purevote.tex
\begin{table}
    \centering
    \begin{tabular}{@{}ll|ll@{}}
    \toprule
    \% Bad sensor & \% Flip-flop & Majority view + reputation & Majority view only \\ \midrule
    1             & 1            & \textbf{0.0002}      & 0.3741                \\
    1             & 4            & \textbf{0.0004}      & 0.3770                \\
    2.5           & 2.5          & \textbf{0.0002}      & 0.3876                \\
    4             & 1            & \textbf{0.0002}      & 0.3701                \\
    10            & 10           & \textbf{0.0023}      & 0.3798                \\ 
    \bottomrule
    \end{tabular}
    \vspace{1mm}
    \caption{False negative \% for majority view plus reputation (\sysname design) vs. majority view only (pure voting)}
    \label{tab:res-ablation-purevote}
\end{table}

%% file: sections/risk-analysis-new.tex
\section{A Model for Risk Analysis}
\label{sec:risk-analysis}

In this section, we introduce a theoretical model of our system, to address two key limitations in our experiments:

\begin{itemize}[leftmargin=*]
    \item \textbf{Extended time span analysis}: our experiments are constrained by computational and storage requirements, limiting their duration. The theoretical model allows us to explore \sysname's long term performance and behavior.
    \item \textbf{Low-probability events}: real-world empirical data reveal that sensor fault rates are small, making it impractical to run long enough simulations for statistically significant results.
\end{itemize}

We analyze the probability that a vehicle's controller would make the wrong decision as a result of receiving erroneous data from neighbors, then confirm consistency with our experiments.

\subsection{Misbehavior probabilities}
\label{sec:misbehavior-probabilities}

Let $P_{false}$ be the probability that a sensor system is not reporting accurate information. 
We start by estimating the value of $P_{false}$ using real-world data, excluding malicious intent due to a lack of data. 
The perception system may be reporting inaccurate information either because of a faulty sensor, or because of faults in some other components of the system (\eg software failure). 
We denote the former as $P_{FS}$ and the latter as $P_{OC}$, where both probabilities are computed when the vehicle is still trusted (bad sensor data may still be accepted). Empirical data suggests a 10\% failure rate for LiDAR sensors over a 10-year period \cite{6644203}, and a vehicle with a faulty sensor is marked untrusted after a conservatively estimated duration, \( T_f \), of 250 seconds.\footnote{Based on our experimental evaluation on the Berlin dataset} This leads us to compute $P_{FS}$ as:\footnote{This represents the proportion of time that a sensor will be trusted while being faulty during a 10-year period}

\revisetwo{
\begin{align}
    P_{FS} = 0.1 \cdot \frac{250}{10 \cdot 365 \cdot 24 \cdot 3600} = 7.9 \times 10^{-8}. 
\end{align}
}

A vehicle may broadcast false information to its neighbors due to reasons other than sensor failures, such as software failure, GPS  failure, camera or radar failure. Based on \cite{BAI2005103}, \cite{GPSFailure}, \cite{6644203}, and \cite{RadarFailure}, and a calculation similar to that for $P_{FS}$, we calculated a value for $P_{OC}$ to be \( 2.79 \times 10^{-7} \).
Combined:
\begin{align}
\label{pfalse_value}
P_{false} = P_{FS} + P_{OC} = 3.59 \times 10^{-7}. 
\end{align}

\subsection{Probability of making wrong decision}
\label{subsec:prob_wrong_decision}
A wrong driving decision is defined as a case in which the final decision of the target vehicle using \sysname differs from the final decision of the target vehicle in an ideal case that all the neighboring faulty sensor vehicles can be verified instantly. 

Referring back to \sysname rules, a target vehicle is exposed to making a wrong decision using if either (i) there is only one trusted neighbor, which sends bad messages before it becomes banned, or (ii) there are $m \ge 2$ trusted neighbors, with at least $\left\lfloor \frac{m}{2} + 1 \right\rfloor$ sending bad messages to the target.
Considering the two cases stated above, the probability of making a wrong decision with $m$ trusted neighbors can be computed by summing over the cases that at least $\left\lfloor \frac{m}{2} + 1 \right\rfloor$ neighbors send false sensing information to the target vehicle:
\begin{multline}  
\label{eqn:eq3}
P_{WD}^{trusted}(m, P_{false}) = 
\sum_{k=\left\lfloor \frac{m}{2} + 1 \right\rfloor}^{m} {m \choose k}(P_{false})^k \cdot \\ (1 - P_{false})^{m-k}.
\end{multline}
We now include both trusted and untrusted vehicles in our analysis, deriving a formula that calculates the probability of erroneous decisions based on the total number of neighbors (as real-world datasets provide data on the density of all vehicles). To do so, we first estimate the probability of being trusted or not, which we denote by $P_{trusted}$ and $P_{untrusted}$, respectively.
Assuming these probabilities are known, the following equation depicts the probability of making a wrong decision when there are $n$ neighbors out of which $m$ are trusted:
\begin{multline}  
\label{eqn:eq4}
P_{WD}^{all}(n, P_{trusted}, P_{false}, T) = 
\sum_{m=T}^{n} {n \choose m}(P_{trusted})^m \cdot \\ (1-P_{trusted})^{n-m}
\cdot P_{WD}^{trusted}(m, P_{false}),
\end{multline}
where $T$ represents the minimum number of trusted neighbors necessary to use the cooperative perception protocol. 

\begin{figure}[tp]
    \centering    
    \includegraphics[width=0.75\columnwidth]{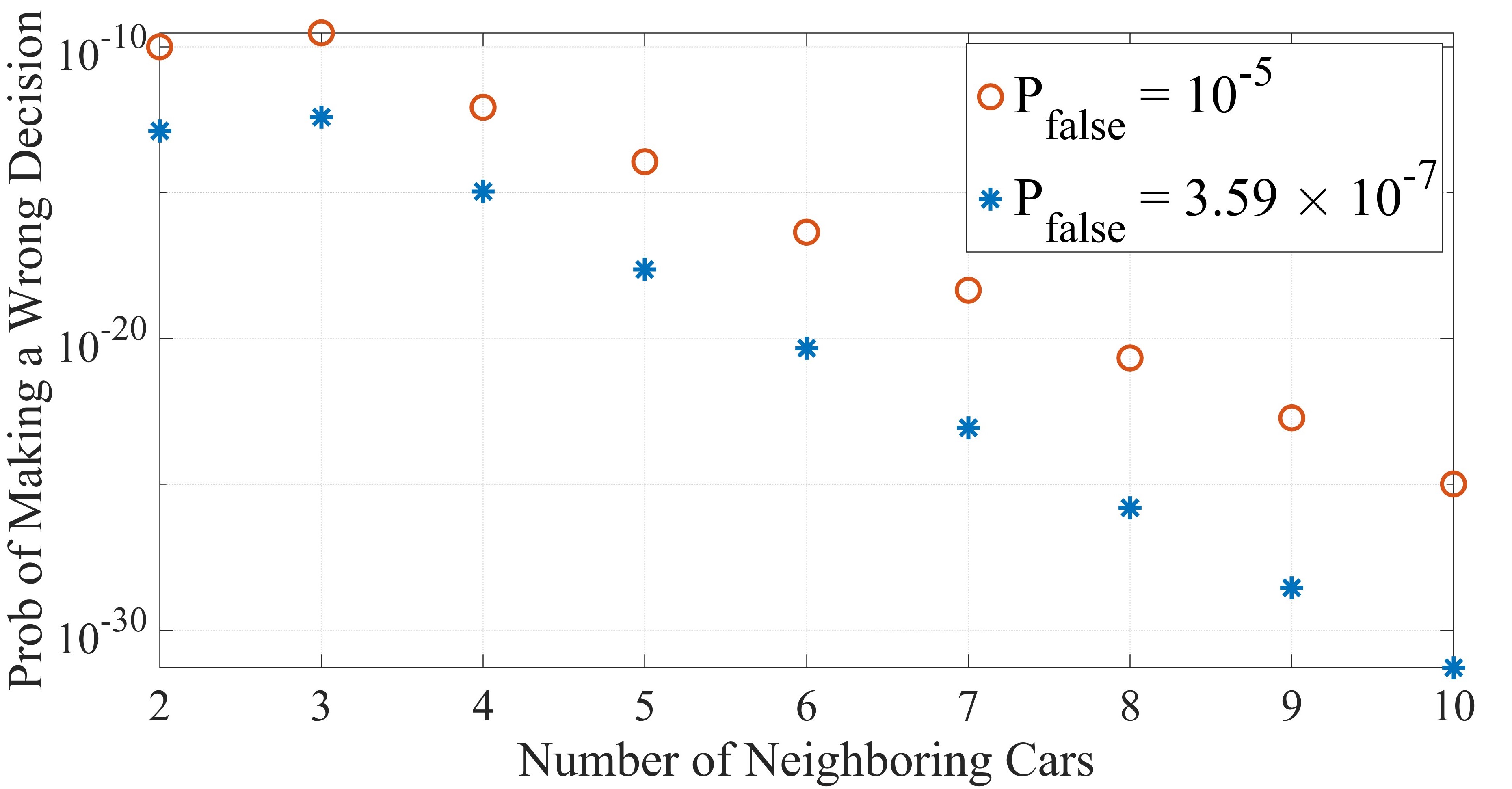}
    \caption{$P_{WD}$ vs. total number of neighbors.}
    \label{fig:prob_making_wrong_decision_vs_all_vehicles}
\end{figure}

Referring to \secref{sec:systemdesign} and \secref{sec:experimental-evaluation}, a vehicle is untrusted and banned in a matter of seconds after a sensor failure. The time it takes for a vehicle to be re-certified varies between hours (authority re-certification) and days (self re-certification). For the purposes of this analysis, we assume an expected time for a vehicle to be re-certified at 7 days (as discussed in \apdxref{appendix:param-selection}).

Disregarding the rare case of a vehicle wrongfully downvoted by a malicious actor, which we omit for simplicity, we calculate the probability of a vehicle being untrusted as the product of the probability of sensor system failure during the lifespan of the sensor system\footnote{This is calculated by the sum of the failure rates for LiDAR, software, GPS, camera, and radar sensors mentioned earlier.},
times the proportion of time during this period that the vehicle will remain banned:

\revisetwo{
\begin{equation}
\label{eqn:eq5}
    P_{untrusted} =0.452 \cdot \frac{7}{10\cdot 365} = 0.0007.
\end{equation}
}

We can now use this value in Eq. \eqref{eqn:eq4} to compute the probability of making a wrong decision given the number of total neighbors.
\figref{fig:prob_making_wrong_decision_vs_all_vehicles} illustrates the corresponding values. The probability of making a wrong decision is insignificant even when the number of trusted neighbors are small.

Last, we derive a formula which computes the probability of making a wrong decision as a function of $P_{false}$, averaged over the number of neighbors $n$ when $m$ of them are trusted:
\begin{multline}
\label{eqn:eq6}
\overline{P_{WD}^{all}}(n, P_{trusted}, P_{false}, T) = \sum_{n=T}^{max\,n} P(n) \cdot \\ P_{WD}^{all}(n, P_{trusted}, P_{false}, T).
\end{multline}
Note that in the above formula, we estimate the distribution of $P(n)$ from the experimental real-world datasets (see \secref{sec:experimental-evaluation}).

\begin{figure}[]
    \centering
    \includegraphics[width=0.70\columnwidth]{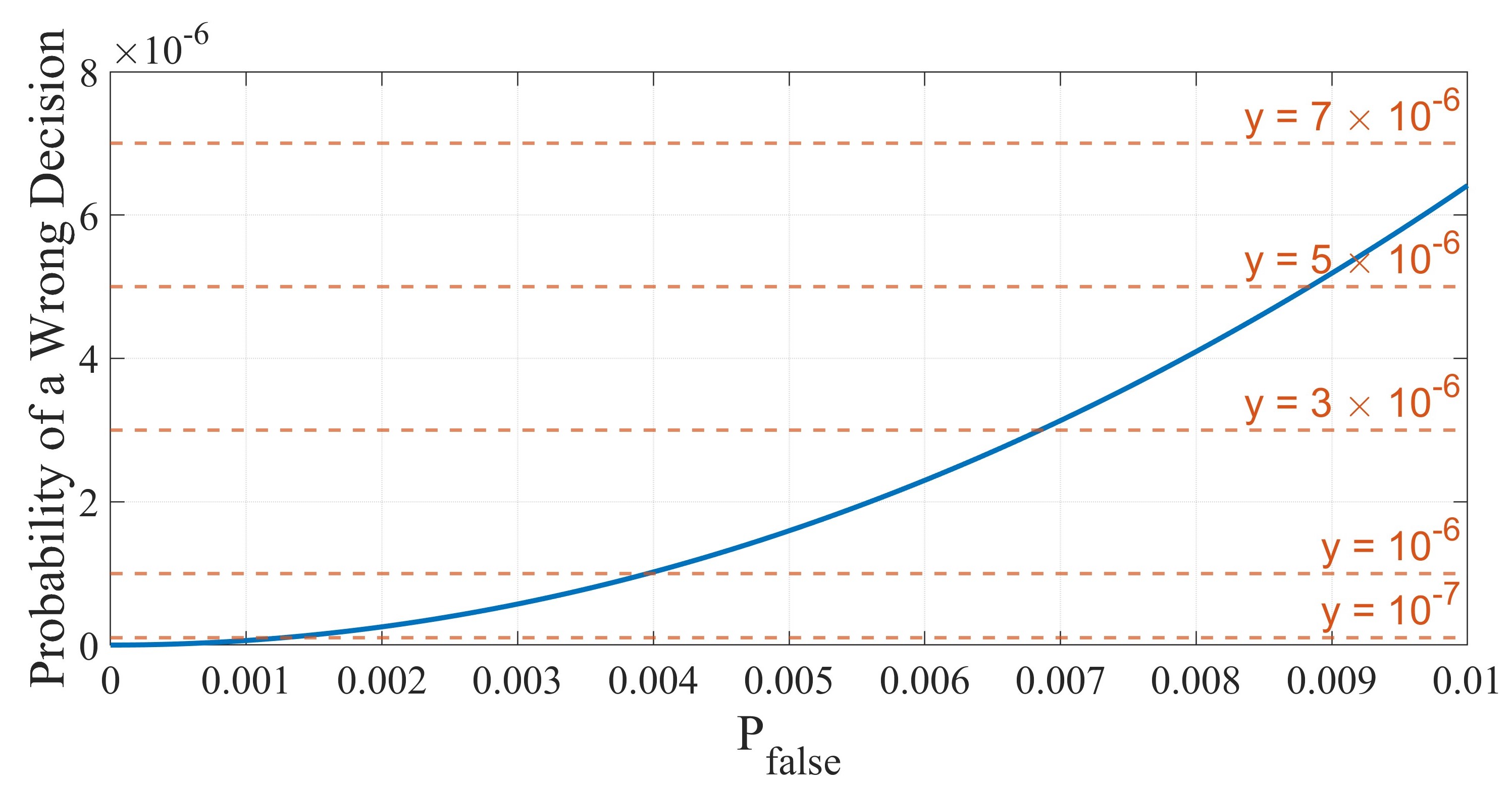}
    \caption{Probability of making a wrong decision vs. $\mathbf{P_{false}}$.}
    \label{fig:prob_making_wrong_decision_vs_Pfalse}
\end{figure}

Figure \ref{fig:prob_making_wrong_decision_vs_Pfalse} plots the probability of making a wrong decision as a function of $P_{false}$, for particularly extreme values, namely $10^{-3}<P_{false}<10^{-2}$, motivated by hypothetical scenarios with a large number of colluding malicious actors.
Even when considering a $P_{false}$ value of 0.01, which substantially exceeds $P_{false}$ as given by Eq. \eqref{pfalse_value}, the probability of an incorrect decision remains low, at less than $7 \times 10^{-6}$.

\subsection{Availability analysis}

We define the availability of the system as the proportion of time during which a vehicle can utilize shared data facilitated by cooperative perception.
Consider a scenario with a single neighbor: If we utilize their data, we increase availability but also risk, as there is no protection should this neighbor sends false data during the short period before it is banned.

To investigate this tradeoff formally, consider there are $n$ neighbors around a target vehicle. Then, the availability of \sysname equals the sum of the probability of scenarios where (i) there is either one or at least two neighbors (depending on whether one opts to use the system in a single neighbor case) and (ii) the majority of the trusted neighbors provide the same information about the environment to the target vehicle.

Given $m$ trusted neighbors, the probability that the data provided by the majority of them are in agreement is $\sum_{k=\left\lfloor \frac{m}{2} + 1 \right\rfloor}^{m} {m \choose k}(1 - P_{false})^k \times P_{false}^{m-k}$.
Thus, the availability $A(n, P_{trusted}, P_{false}, T)$ when there is $n$ total neighbors is:
\begin{multline}  
\label{eqn:eq7}
A(n, P_{trusted}, P_{false}, T) = \sum_{m=T}^{n} {n \choose m}P_{trusted}^m \cdot \\ (1-P_{trusted})^{n-m} 
\cdot \sum_{k=\left\lfloor \frac{m}{2} + 1 \right\rfloor}^{m} {m \choose k}(1 - P_{false})^k \cdot P_{false}^{m-k}.
\end{multline}

To compute the average usability of \sysname we average over the number of neighbors similar to Eq. \eqref{eqn:eq6} and get:
\begin{multline}
\label{eqn:eq11}
\overline{A}(n, P_{trusted}, P_{false}, T) = \sum_{n=1}^{max\,n} P(n) \cdot \\ A(n, P_{trusted}, P_{false}, T).
\end{multline}

We use the values of $P_{false}$ and $P_{trusted}$ mentioned above, and the distribution of the number of neighbors $P(n)$ estimated from the experimental real-world datasets, to 
compute $\overline{A}(n, P_{trusted}, P_{false}, T)$ and $\overline{P_{WD}^{all}}(n, P_{trusted}, P_{false}, T)$ and investigate the availability-risk tradeoff.
As illustrated in Table \ref{tab:aavg}, by allowing data sharing in the presence of a single trusted neighbor, we improve availability. While the probability of making the wrong decision also increases, it is still minuscule (less than $2\times 10^{-8}$), suggesting that \sysname's hybrid solution ($T = 1$) increases availability without affecting practical safety.

\begin{table}[!htp]
    \centering
        \begin{tabular}{@{}lll@{}}
        \toprule
        & $\overline{A}$   & \shortstack{$\overline{P_{WD}^{all}}$} \\ \midrule
        \begin{tabular}[c]{@{}l@{}}2+ vehicles \end{tabular} & 0.9504 & $3.1522 \times 10^{-14}$                 \\
        \begin{tabular}[c]{@{}l@{}}1 vehicle OK\end{tabular}           & 0.9999 & $1.7011 \times 10^{-8}$                 \\ \bottomrule
        \end{tabular}
    \caption{Availability vs. $P_{WD}$ w.r.t. minimum majority}
    \label{tab:aavg}
\end{table}
\vspace{-1em}

\begin{figure}
    \centering
    \includegraphics[width=0.70\columnwidth]{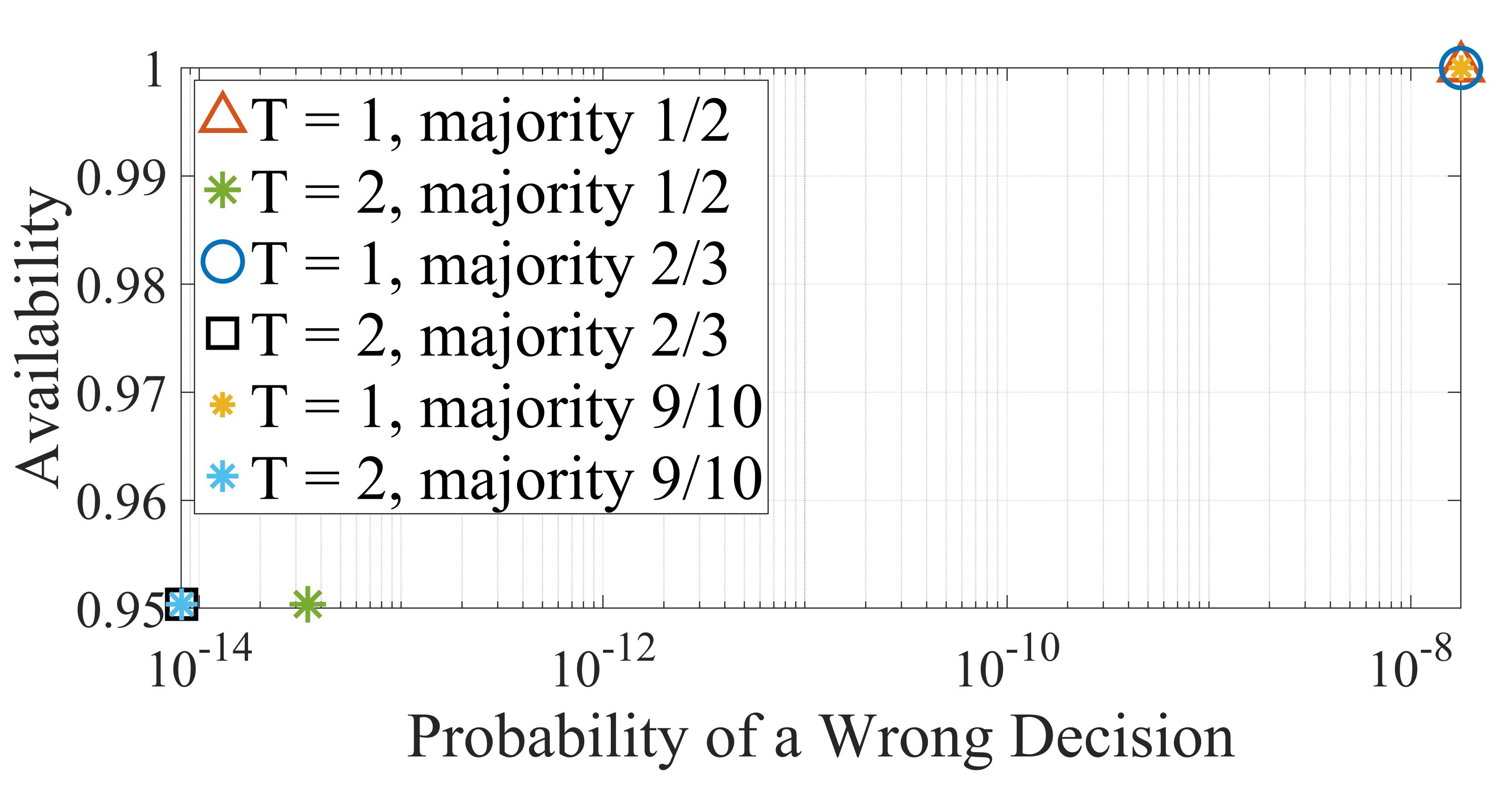}
    \caption{Availability-risk tradeoff. (Majority ratios represent the minimum fraction of shared data required to be consistent.)}
    \label{fig:availability_vs_probwrongdecision}
\end{figure}

Last, Figure \ref{fig:availability_vs_probwrongdecision} presents the availability-risk tradeoff for a number of parameters, supporting our approach of $T = 1$. Intuitively, this is because once a vehicle fails, it will quickly become untrusted, and unless multiple attackers are colluding, it is impossible in practice for more than one vehicle to happen to be at the same space and time in this undetected phase.

\subsection{Model-experiment correspondence}

\input{tables/model-vs-experiment}

We use Eq. \eqref{eqn:eq6} with
$P_{false}$ and $P_{trusted}$ estimated from the experimental evaluation to compare the model with experimental results under the same conditions. Comparative analysis of false negative message percentages presented in \tabref{tab:model_vs_experiment} illustrates the concordance between our theoretical model and the empirical data, affirming the model’s accuracy in mirroring real-world dynamics.

While the model's false negative message percentages and the empirical figures generally agree, discrepancies are noted at higher probability values, where the model predicts higher percentages. This variation is primarily due to the conservative assumptions in the model, which incorporate an upper bound for the time it takes to ban a vehicle disseminating incorrect information. In actual scenarios, the banning process is often expedited, resulting in a reduced false negative percentage. We postulate that if the simulations were to extend beyond the 700-second mark, the model's predictions would exhibit an even greater degree of convergence with the observed data.

The observed pattern of consistency (notable when considering the model's conservativeness) lends credibility to the model and supports its utility for long-term applications. It showcases the model's capacity as a robust predictive tool, capable of effectively forecasting system behavior over extended periods.

%% file: tables/model-vs-experiment.tex
\begin{table}[!htp]
  \centering
  \small
  \begin{tabular}{ll|l|l}
    \toprule
   \multirow{2}{*}{\shortstack{\% Bad\\sensor}} & \multirow{2}{*}{\shortstack{\% Flip-\\flop}} &  \multirow{2}{*}{\shortstack{Experimental\\false negative \%}} &
    \multirow{2}{*}{\shortstack{Risk model\\false negative \%}}\\
          &       &     &      \\ \midrule
          
    0.1 & 0.1 & 0.0001 & 0.0001    \\
    1   & 1   & 0.0055 & 0.0048    \\
    1   & 4   & 0.0069 & 0.0058    \\
    2.5 & 2.5 & 0.0051 & 0.0060    \\
    4   & 1   & 0.0038 & 0.0087    \\
    \bottomrule
  \end{tabular}
  \caption{False negative message percentages as calculated by our risk model vs. measured from our experiments.}
  \label{tab:model_vs_experiment}
\end{table}

%% file: sections/related-works.tex
\section{Related Work}

V2X communications for cooperative autonomy thus far focuses on either direct broadcast \cite{AVR,autocast} or infrastructure-assisted (vehicle to infrastructure to other vehicles) \cite{emp,chen2019fcooper}. \cite{securingv2x} describes 3 primary threats to such a V2X system: packet flooding, Sybil nodes, and false information.

\cite{vanetjam,ai-jam-detect} can detect DoS/DDoS type attacks by detecting abnormalities in the physical layer of the V2X connection; however, an effective defense against all possible types of flooding can be difficult in a wireless V2X environment.

\cite{ifal,ETSI_ITS_TS_103_097,autosec2022_tee} prevents Sybil nodes by utilizing a trusted, centralized authority to assign identities to vehicles through the use of PKI certificates with pseudonyms to alleviate privacy concerns. \cite{mixzones,changing-certs-no-good,BSPpseudonym} discusses strategies for pseudonym-switching. \cite{ifal} provides efficient, wide-scale revocation of certificates by requiring that pseudonyms be continually activated by the use of a short code (enabling CRL-less revocations). \cite{autosec2022_tee} goes even further and uses a TEE to additionally provide confidentiality, preventing untrusted entities from reading V2X data as it is encrypted. These systems, however, rely on manual response, raising the question we have discussed earlier in \secref{sec:introduction}.

\cite{Schmidt2008VehicleBA,TFDD} proposes a system where vehicles in the V2X network can evaluate the reliability of others, locally compute trust, then share results with a global list \revise{(introducing the concept of \textit{node based trust})}. However, there is weak protection against Sybil nodes, and pseudonyms are not supported.

\cite{VanetReputation} proposes a reputation-based system to filter out untrustworthy peers, with support for pseudonyms and trusted distribution of reputation scores. However, being a pure reputation based system that only allows one vote per epoch, there is a larger delay in responding to misbehaving peers.

Expanding on misbehavior detection, \cite{plausibilitycheck} introduces a low-overhead, local consistency checking algorithm \revise{following the \textit{data based trust} concept} with particle filters. \cite{CAMPMisbehaviorDetection} evaluated a similar system for local misbehavior detection based on proximity plausibility in the real world, \revise{\cite{CADConsistencyChecking} goes further and both introduces and evaluates real-world attacks and defense algorithms for data validation, showing that it is possible to effectively implement these building blocks.}

\cite{TruPercept} provides a similar visibility-based checker and even a global trust aggregation system. However, it has high overhead and a short look-back window, resulting in the system \textit{forgetting} past misbehaviors in a few minutes.

%% file: sections/conclusion.tex
\section{\space \space Limitations \& Future Work}

\parab{Majority view protocol limits.}
While \sysname can utilize in-situ majority view aggregation to filter out bad messages even from a trusted vehicle, it can only do so if there are enough vehicles ($n \ge 3$) sharing co-visibility of the same objects nearby. Otherwise, vehicles will fall back to a reputation-only system. We \revise{believe further improvements are possible by} integrating other data validation methods (\eg kinematic) into the pipeline.

\parab{\revise{Broadcast V2X assumptions.}}
\revise{\sysname assumes that V2X messages are broadcasted and will be received by anyone within range. Should a malicious actor be able to single out specific vehicles to attack (\ie with a directional antenna), \sysname will still be able to mitigate wide-scale attacks, but may not be able to completely mitigate an attack against a single vehicle. Again, we believe improvements are possible by integrating other data validation methods into the pipeline.}

\parab{Centralization.}
A central authority (the SA) is still required in order to process votes, map pseudonymous identities to actual identities, and enforce security policies. This presents a security/privacy risk if the SA is breached, or if the SA itself decides to collect data from peers. \revise{In this area, recent works such as \cite{RTracing,SuBlockchain,PPBlockchain} have proposed utilizing decentralized methods, such as blockchain ledgers, to help solve this issue.}

\section{\space \space \space Conclusion}

We have designed, implemented, formally analyzed, and experimented with \sysname, a cooperative autonomy message filtering system that uniquely integrates the strengths of reputation-based and majority-based methodologies, ensuring the reliability and integrity of cooperative perception data.

Our city-scale {simulations} validates the effectiveness of \sysname under realistic traffic scenarios, allowing the vast majority of vehicles to benefit from cooperative perception while swiftly dealing with misbehaving vehicles, preventing bad messages from influencing vehicles' control decisions.

Finally, we developed a detailed mathematical model for \sysname, which further demonstrates the system's robustness and reliability over extended periods, under various conditions. 

%% file: sections/acknowledgement.tex
\section*{Acknowledgement}

We would like to thank J. Aydin, K. Liu, M. Sperling, and S. A. Choe for their contribution of code to the \sysname simulation environment. 

Additionally, we thank the anonymous reviewers for their insightful suggestions and feedback during the review process.

This material is based upon work supported by the National Science Foundation under award CNS-1956445.

%% file: sections/appendix-parameter-selection.tex
\section{Parameter derivation}
\label{appendix:param-selection}

\subsection{Inter-vote epoch ($T_{IVE}$)}
\label{appsec:tive}

$T_{IVE}$ controls how often a single vehicle $A$ can perform voting actions for another vehicle $B$ in a specific time period, ensuring that reputation-affecting votes comes from as diverse viewpoints (that is, multiple vehicles) as possible.

Since the SA can measure the distribution of the \textit{inter-accepted votes interval (IAVI)}, the 95th percentile IAVI can be used along with expected travel times \cite{CensusCommuteTime} to calculate the expected votes received by a vehicle per day. Then, a value for recovery time (from lowest trust score, to fully trusted) can be calculated. Following the argument that every one of these votes should ideally be unique, this recovery time value can then be used as the updated $T_{IVE}$. The SA could continually run this algorithm to automatically tune $T_{IVE}$.

As an example based on our datasets, we found that we have the 95th percentile IAVI at about half a minute, resulting in the expected number of votes that a vehicle would get in a day in the order of a few hundred. Extrapolated, this amounts to approximately a week, which can be a sensible default value.

\subsection{Inter-downvote epoch ($T_{IDE}$)}
\label{appsec:tide}

The optimal value for the inter-downvote epoch $T_{IDE}$ is a balance between the effectiveness of \sysname's downvoting mechanism and the risk for denial-of-service exploits.

To derive \revise{the global base value for} $T_{IDE}$, the SA first defines the probability of voting eligibility ($P_{VE}$) as the probability that a vehicle has not previously downvoted any vehicle, and is therefore eligible to vote. This can be expressed as
$P_{VE} = 1 - P(X \text{ downvoted at least one vehicle within interval } T_{IDE})$.

We compute $P_{VE}$ by determining the ratio of vehicles that have engaged in downvoting to the total number of vehicles:
Let $C_d$ represent the number of vehicles that have downvoted at least one vehicle within an interval $T_{IDE}$, and let $C_t$ denote the total number of vehicles in the area on that day. Therefore:

\begin{equation}
    \label{eqn:eq_Appendix_1}
    P_{VE} = 1 - \frac{C_d}{C_t} .
\end{equation}

Following the definition of $P_{VE}$ in Eq. \refeq{eqn:eq_Appendix_1}, we now derive the probability that a given untrusted vehicle can be successfully downvoted by its neighbors (denoted as $P_{DS}$), as follows:
\begin{multline}  
\label{eqn:eq_Appendix_2}
P_{DS} = \sum_{n}^{} P(n) \cdot (\sum_{k=1}^{n} {n \choose k} \cdot P_{VE}^k \cdot (1 - P_{VE})^{n-k})  .
\end{multline}

In the above formula, \( P(n) \) represents the probability of having \( n \) neighbors. For each possible \( n \), we accumulate the probabilities that at least one of these peers is eligible to vote. 

The SA can use an online learning approach to continuously update \revise{base} $T_{IDE}$, by measuring values to calculate \eqnref{eqn:eq_Appendix_1} and \eqnref{eqn:eq_Appendix_2}, in order to optimize $P_{DS}$ to the desired risk threshold.  In our experiments we assumed a risk threshold of $P_{DS} = 0.99$ for a given $T_{IDE}$ value. 

\subsection{Vote submission time limit ($T_{vote}$)}
\label{appsec:tvote}

The optimal value of $T_{vote}$ follows the consideration of a bound in which $T_{vote} \ge T_{BBI}$, the later representing the beacon broadcast interval discussed later in the appendix (as otherwise there may be gaps in which voting cannot occur, due to how \sysname process votes). However, a large value for $T_{vote}$ may result in increased vulnerability to replay attacks.

Therefore, to automatically learn $T_{vote}$, an online learning approach can be used, where we define $T_{vote} = T_{BBI} + \delta$, where $\delta$ is the 95th percentile measured voting network latency, guaranteeing that the majority of valid voting messages will be preserved.
{Based on real-world measurements of cellular network latency \cite{4GLatency}, we used $\delta = 50$ ms for our experiments.}

\subsection{Timeout interval ($T_{TI}$)}
\label{appsec:tti}

The timeout interval $T_{TI}$ controls, on a per-vehicle level, how long \textit{yellow} and \textit{red} flags stay on that specific vehicle. (This is an adaptive value that gets multiplicatively increased with every \textit{red} flag that a vehicle gets, as we consider vehicles to be less trustworthy if they have been previously banned.)

As $T_{TI}$ also controls how fast a \textit{yellow} flag (which does not ban a vehicle) gets cleared, starting out with too low of a value may enable an attacker to exploit this window to perform one attack every $T_{TI}$ period without getting banned, while starting out with a high value results in reduced availability.

Therefore, an initial value of $T_{TI}$ at 7 days may be a good starting point, to assure that vehicle owners will have at least one weekend to perform maintenance if their vehicle was banned, as well as lower the risk of an attack by limiting total opportunity window available for crafty attackers
{(\eg with an initial value of $T_{TI} = 7$ days, an attacker can attack the system at most $T_{FW}$ every week, which in the default case of $T_{FW} = 20$s, is only $0.0032\%$ of the time).}

\subsection{Flagging window duration ($T_{FW}$)}
\label{appsec:flagging-window-duration}

The length of flagging window slots, $T_{FW}$, controls the level of certainty for misbehavior before a vehicle is banned.

{A long $T_{FW}$ can make sure that downvotes to a vehicle comes from multiple events, increasing the certainty that a vehicle is truly misbehaving, before it gets banned. This, however, results in a longer time to ban a misbehaving vehicle, allowing a \textit{flip-flopping} attacker to potentially perform more attacks before it gets completely excluded from the network.}

{However, if $T_{FW}$ is set too low, \sysname can mistakenly split downvotes from the same event into multiple ones, resulting in potential nuisance bans and easier denial-of-service attacks.}

We believe that a reasonable value would be in the tens of seconds and performed empirical simulations on the Boston dataset with different values of $T_{FW}$. Based on our results in \tabref{tab:apdx-trf}, we found $T_{FW} = 20$ to yield acceptable performance.

We envision that the SA would be able to monitor, offline, the percentage of vehicles with faulty sensors or potential attackers, as well as traffic patterns and statistics (either from counting, or V2X methods such as \cite{InfraLessDensityEstimation}), then utilize ground-truth simulations of \sysname with the aforementioned data to calculate and tune $T_{FW}$ following our derivation here.

\input{tables/trf-tables.tex}

\subsection{Beacon broadcast interval ($T_{BBI}$)}
\label{appsec:tbbi}

For $T_{BBI}$, we consider the balance between maximizing airtime efficiency on the control channel, and minimizing the time a vehicle needs in order to "get to know their neighbor".

Assuming a V2X range of 400m \cite{QualcommCV2X}, and a maximum traffic speed of 136 km/h (maximum limit in the U.S.), we estimate that two vehicles will be in range 5.2 seconds before meeting.

With an average beacon size of $1300$ bytes, we can estimate the bandwidth requirements of \sysname's beacons, assuming that the channel is time-division multiplexed and collision-free. {With typical channel capacities between 3 and 10 Mbps \cite{LTE-Direct,80211p}, this yields around 280 to 960 slots per second.}

Based on analysis of our datasets, we found that within the radio range of 400 meters, the maximum number of vehicles in radio contact with each other does not exceed 240, {bounding the ideal transmission rate as a range between $[1, 4]$ Hz}.

Allowing for headroom, we suggest $T_{BBI} = 1 Hz$ to give vehicles enough time to be able to receive beacons and still have time to react while keeping bandwidth requirements low.

\subsection{Trust score threshold ($N_{thresh}$)}
\label{appsec:nthresh}

$N_{thresh}$ controls the SA-internal threshold for the trust score that transitions vehicles from trusted to untrusted state without a ban, enabling \sysname to quickly react to potential misbehavior, while maintaining robustness against false votes.

We performed empirical studies that shows that $N_{thresh}$ should be such that 2 other vehicles' downvotes are required before a vehicle will become untrusted, by running \sysname with different $N_{thresh}$ values corresponding to different number of downvoting vehicles $N_{D}$ required for temporary exclusion.

Our results in \tabref{tab:res-nthresh-comparison} shows an increased false negative message percentage for every additional downvoter requirement, which follows the logic that the higher the required number of downvoters (lower $N_{thresh}$), the lower the performance for quick response \revise{as meeting the requisite number of downvoting vehicles in that short time span becomes more challenging. As there are no improvements in false positives for $N_{D} > 2$, a threshold as presented could be a sensible default.}

\input{tables/nthresh-downvoters.tex}

%% file: tables/trf-tables.tex
\begin{table}[]
\begin{tabular}{ll|llllllll}
\hline
\multicolumn{1}{c}{\multirow{3}{*}{\shortstack{\% Bad\\sensor}}} & \multicolumn{1}{c|}{\multirow{3}{*}{\shortstack{\% Flip\\flop}}} & \multicolumn{4}{c|}{\textbf{False negative \%}} & \multicolumn{4}{c}{\textbf{False positive \%}} \\ \cline{3-10} 
\multicolumn{1}{c}{} & \multicolumn{1}{c|}{} & \multicolumn{8}{c}{$T_{FW}$} \\ \cline{3-10} 
\multicolumn{1}{c}{} & \multicolumn{1}{c|}{} & \multicolumn{1}{c}{\textbf{10}} & \multicolumn{1}{c}{\textbf{20}} & \multicolumn{1}{c}{\textbf{40}} & \multicolumn{1}{c|}{\textbf{80}} & \multicolumn{1}{c}{\textbf{10}} & \multicolumn{1}{c}{\textbf{20}} & \multicolumn{1}{c}{\textbf{40}} & \multicolumn{1}{c}{\textbf{80}} \\ \hline
1 & 1 & 0.05 & 0.03 & 0.03 & \multicolumn{1}{l|}{0.04} & 0.26 & 0.26 & 0.26 & 0.25 \\
2.5 & 2.5 & 0.02 & 0.02 & 0.02 & \multicolumn{1}{l|}{0.03} & 0.65 & 0.64 & 0.65 & 0.63 \\
1 & 4 & 0.01 & 0.02 & 0.02 & \multicolumn{1}{l|}{0.02} & 0.43 & 0.42 & 0.42 & 0.41 \\
4 & 1 & 0.03 & 0.03 & 0.05 & \multicolumn{1}{l|}{0.05} & 1.13 & 1.13 & 1.13 & 1.12 \\
10 & 10 & 0.07 & 0.12 & 0.17 & \multicolumn{1}{l|}{0.31} & 2.98 & 2.97 & 2.93 & 2.88 \\ \bottomrule
\end{tabular}
\caption{FN/FP \% vs. $T_{FW}$ changes.}
    \label{tab:apdx-trf}
\end{table}

%% file: tables/nthresh-downvoters.tex
    \begin{table}

    \begin{tabular}{@{}ll|lll|lll@{}}
        \toprule
        \multirow{2}{*}{\shortstack{\% Bad\\Sensor}} & \multirow{2}{*}{\shortstack{\% Flip-\\flop}} & \multicolumn{3}{l|}{False negative \% with $N_{D}$} & \multicolumn{3}{l}{False positive \% with $N_{D}$} \\ \cmidrule(l){3-8} 
         &  & $1$ & $2$ & $3$ & $1$ & $2$ & $3$ \\ \midrule
        1 & 1     & 0.016 & 0.043 & 0.080 & 0.295 & 0.290 & 0.290 \\
        2.5 & 2.5 & 0.027 & 0.055 & 0.087 & 0.692 & 0.679 & 0.679 \\
        1 & 4     & 0.021 & 0.075 & 0.136 & 0.496 & 0.490 & 0.490 \\
        4 & 1     & 0.026 & 0.052 & 0.113 & 1.053 & 1.030 & 1.030 \\
        10 & 10   & 0.110 & 0.266 & 0.408 & 2.864 & 2.812 & 2.811 \\ \bottomrule
        \end{tabular}
        \vspace{1mm}
        \caption{Comparison results for different downvoter thresholds.}
        \label{tab:res-nthresh-comparison}
    \end{table}